\documentclass{article}

\PassOptionsToPackage{numbers, compress}{natbib}

\usepackage[final]{neurips_2025}

\usepackage[utf8]{inputenc} 
\usepackage[T1]{fontenc}    
\usepackage{hyperref}       
\usepackage{url}            
\usepackage{booktabs}       
\usepackage{amsfonts}       
\usepackage{amsmath}
\usepackage{graphicx}
\usepackage{multirow}
\usepackage{nicefrac}       
\usepackage{microtype}      
\usepackage{xcolor}         
\usepackage{subcaption}
\usepackage{caption}
\usepackage{cleveref}

\title{BRACE: A \underline{B}enchmark for \underline{R}obust \underline{A}udio \underline{C}aption Quality \underline{E}valuation}

\author{
Tianyu Guo\thanks{Equal contribution.} \\
Peking University \\
\And
Hongyu Chen\footnotemark[1] \\
Peking University \\
\And
Hao Liang\footnotemark[1]\,\,\thanks{Project Leader.} \\
Peking University \\
\And
Meiyi Qiang \\
Peking University \\
\And
Bohan Zeng \\
Peking University \\
\And
Linzhuang Sun \\
University of Chinese Academy of Sciences \\
\And
Bin Cui\thanks{Corresponding Author.} \\
Peking University \\
\And
Wentao Zhang\footnotemark[3] \\
Peking University \\
}

\begin{document}

\maketitle

\begin{abstract}

  Automatic audio captioning is essential for audio understanding, enabling applications such as accessibility and content indexing. However, evaluating the quality of audio captions remains a major challenge, especially in reference-free settings where high-quality ground-truth captions are unavailable. While CLAPScore is currently the most widely used reference-free Audio Caption Evaluation Metric(ACEM), its robustness under diverse conditions has not been systematically validated.
  To address this gap, we introduce BRACE, a new benchmark designed to evaluate audio caption alignment quality in a reference-free setting. BRACE is primarily designed for assessing ACEMs, and can also be extended to measure the modality alignment abilities of Large Audio Language Model(LALM). BRACE consists of two sub-benchmarks: BRACE-Main for fine-grained caption comparison and BRACE-Hallucination for detecting subtle hallucinated content. We construct these datasets through high-quality filtering, LLM-based corruption, and human annotation. 
  Given the widespread adoption of CLAPScore as a reference-free ACEM and the increasing application of LALMs in audio-language tasks, we evaluate both approaches using the BRACE benchmark, testing CLAPScore across various CLAP model variants and assessing multiple LALMs.
  Notably, even the best-performing CLAP-based ACEM achieves only a 70.01 F1-score on the BRACE-Main benchmark, while the best LALM reaches just 63.19. 
  By revealing the limitations of CLAP models and LALMs, our BRACE benchmark offers valuable insights into the direction of future research. Our evaluation code and benchmark dataset are released in \url{https://github.com/HychTus/BRACE_Evaluation} and \url{https://huggingface.co/datasets/gtysssp/audio_benchmarks}.
  
\end{abstract}

\section{Introduction}\label{sec: introduction}
Recently, audio captioning data has gained increasing importance in multimedia understanding and accessibility, as it enables the effective interpretation of audio content through textual descriptions. This emerging field is essential for applications such as content indexing, searchability, and providing accessibility to users with hearing impairments.

A few pioneering works have contributed to the development of audio benchmarks for evaluating the performance of audio language models. FENSE Benchmark~\cite{Zhou2021CanAC} is one such benchmark, designed for pairwise comparison of audio caption quality. Comp-A~\cite{ghosh2023compa} was developed to assess whether an audio language model can accurately understand the order and occurrence of acoustic events in audio. Furthermore, audio hallucination detection~\cite{kuan2024understanding, kuan2024can} evaluates the hallucination tendencies of audio language models by directly altering the entities in a caption.
\begin{figure}
  \centering
  \includegraphics[width=0.8\textwidth]{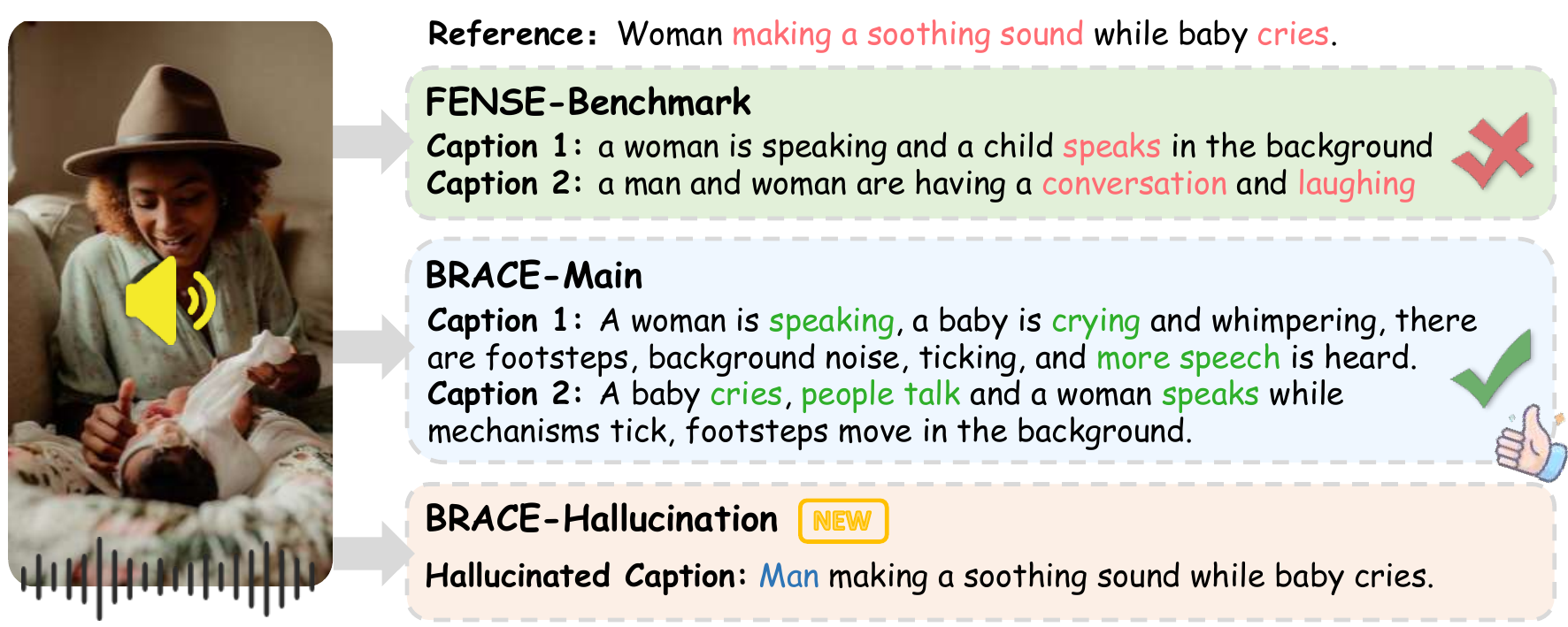}
  \caption{We present an example from our BRACE benchmark, which offers greater detail compared to the FENSE Benchmark. Additionally, our benchmark includes audio-caption pairs, whereas FENSE only contains captions for audio. Furthermore, we introduce a new Hallucination benchmark, named BRACE-Hallucination, as shown at the bottom, for detecting audio-caption hallucinations.}
  \label{fig: Face_1}
  \vspace{-6mm}
\end{figure}

However, previous research has not constructed a robust benchmark for evaluating reference-free ACEMs, nor for detecting whether such metrics can identify object-based hallucinated data in audio captions.

\paragraph{Lack of Audio-Caption Quality Evaluation Benchmark}  
FENSE~\cite{Zhou2021CanAC} proposed a benchmark relying solely on audio captions for pairwise comparison, failing to fully leverage the information from the audio modality. Additionally, the models used in FENSE Benchmark~\cite{Zhou2021CanAC} for caption generation are outdated. More recent models, such as LTU~\cite{Gong2023ListenTA} and GAMA~\cite{Ghosh2024GAMAAL}, can generate human-like captions, which make pairwise comparison tasks significantly more challenging. Comp-A~\cite{ghosh2023compa} primarily focuses on the temporal aspects of audio, highlighting the need for a benchmark that evaluates audio caption quality.

\paragraph{Lack of High-Quality Hallucination Benchmark}  
In the era of large multimodal models(LMM), hallucination detection has become a critical component for evaluating LALMs. Although previous studies~\cite{kuan2024understanding, kuan2024can} have proposed methods for detecting audio hallucinations, these approaches typically involve simple questions such as, ``Can you detect the sound of a \textbf{dog (true)} in the audio?" or ``Can you detect the sound of a \textbf{cat (hallucination)} in the audio?". This type of questioning helps the model identify the exact locations of hallucinations within the caption, allowing it to focus on specific terms rather than making a judgment about the overall content of the caption. This approach reduces the difficulty for the model in detecting hallucinations in the caption.
In practice, hallucinations in language model outputs cannot always be detected in such a direct manner. Therefore, a more comprehensive hallucination benchmark, featuring full-length audio-caption pairs, is required.


\paragraph{Importance of Reference-free Audio Caption Evaluation Metrics}
The reference-based evaluation method for audio captioning depends on the availability of high-quality reference captions. However, compared to speech data \cite{panayotov2015librispeech}, the amount of high quality audio-caption data is relatively limited. Currently, commonly used datasets like AudioCaps \cite{kim2019audiocaps} and Clotho \cite{drossos2020clotho} contain only about 45,000 audio samples in total, with an approximate total audio duration of 150 hours. Moreover, in practice, we have found that the quality of multiple captions for many audio samples varies significantly. This leads to situations where reference-based methods may sample low-quality reference captions. Therefore, a robust reference-free ACEM becomes particularly important for effectively evaluating large-scale datasets when reference captions are noisy or inconsistent.

To address these issues, we introduce a new benchmark: BRACE. This benchmark consists of two sub-benchmarks. The main benchmark, BRACE-Main, is designed for comparing audio captions and includes three categories: HH, HM, and MM, where ``H" refers to human-annotated captions and ``M" refers to machine-generated captions (i.e. from LALMs). Each audio clip is associated with multiple caption pairs. For each caption pair, CLAP separately evaluates the alignment score between the audio and each caption, selecting the caption that is better aligned with the audio. In contrast, LALM jointly takes the audio and the caption pair as input and directly determines which caption is more consistent with the audio.
To construct the BRACE-Main benchmark, we first filter high-quality audio-caption pairs. Next, we automatically generate and corrupt captions to create additional audio-caption pairs. Then, three experienced annotators evaluate each pair and select the better one, with consensus required for selection. To construct the BRACE-Hallucination benchmark, we utilize large language models to identify nouns within captions and replace them with alternative nouns, ensuring that logical consistency is maintained before and after the substitution. 

Our contributions are summarized as follows:
\begin{itemize}
    \item We developed a new reference-free audio-caption pairwise comparison benchmark, BRACE-Main, specifically designed to evaluate the caption quality evaluation capabilities of CLAP models used in CLAPScore, as well as modality alignment capability of LALMs.
    
    \item We introduce BRACE-Hallucination, a novel benchmark designed to detect subtle hallucinated content, presenting greater challenges for both CLAP models and LALMs, and enabling more rigorous evaluation of their fine-grained audio-text alignment capabilities.
    \item We comprehensively evaluated LALMs and CLAP-based ACEMs on BRACE, revealing their weaknesses and informing future improvements in audio-language understanding.
\end{itemize}

\section{Related Work}

\subsection{Audio Caption Evaluation}

\paragraph{Linguistic Evaluation.} Traditional evaluation methods for audio captioning are adapted from natural language generation (NLG) techniques, based primarily on simple matching between reference and candidate captions. Metrics such as BLEU~\cite{Papineni2002BleuAM} and ROUGE~\cite{Lin2004ROUGEAP} employ N-gram matching. METEOR~\cite{Banerjee2005METEORAA} improves semantic alignment by incorporating synonym matching and stemming, while CIDEr~\cite{Vedantam2014CIDErCI} utilizes TF-IDF weighting to emphasize the importance of key terms. SPICE~\cite{Anderson2016SPICESP}, which focuses on matching object graphs in captions, places greater emphasis on semantic alignment, and SPICEr~\cite{Liu2016ImprovedIC}, a combination of CIDEr and SPICE, aims to balance both syntactic and semantic evaluation.

However, the diversity of potential captions for the same audio, together with the inherent ambiguity of the audio content, increases the variability of the captions~\cite{Bhosale2022TexttoAudioGB}. These factors lead to a low correlation between these simple matching metrics and human judgment~\cite{Zhou2021CanAC}.

\paragraph{Reference Based Evaluation.}  To better assess whether reference and candidate captions alignment, FENSE~\cite{Zhou2021CanAC} employs a pre-trained language model to compute the BERT-Score~\cite{Zhang2019BERTScoreET}. This approach encodes both candidate and reference sentences as vectors and computes the cosine similarity between them as a measure of alignment. This improves the assessment of semantic alignment. 
ACES~\cite{Wijngaard2023ACESEA} improves interpretability by extracting sound descriptors from captions and calculating cosine similarity for fine-grained matching. The s2vscore~\cite{Bhosale2022TexttoAudioGB} generates embeddings for acoustically similar sounds, providing a more accurate assessment of the acoustic consistency of captions rather than their semantic alignment. 


However, all of these metrics are reference-based, meaning that they do not incorporate the original audio into the evaluation process. Instead, they measure the degree of match between candidate and reference captions. These methods are primarily designed to evaluate models trained on reference captions and cannot be applied for broader audio captioning evaluations.

\paragraph{Reference-free Evaluation.} CLAPScore is a recently proposed reference-free evaluation metric. CLAP models~\cite{Elizalde2023CLAPLA, elizalde2023clap, elizalde2024natural, niizumi2024m2d-clap, wu2024largescalecontrastivelanguageaudiopretraining}, trained via contrastive learning, map both audio and text into a shared vector space. We can measure caption quality by calculating the cosine similarity between audio and text embeddings. CLAPScore represents how well the caption aligns with the original audio.

\subsection{Existing Benchmarks for Evaluating Metrics}

FENSE Benchmark~\cite{Zhou2021CanAC} is proposed to assess the effectiveness of audio captioning evaluation metrics. It focuses on the correlation between metric scores and human rankings of captions. However, the models used in FENSE Benchmark~\cite{Zhou2021CanAC} for caption generation are outdated, which limits their relevance in the context of recent advancements in LLMs.


VATEX-EVAL and ActivityNet-FOIL~\cite{Shi2021EMScoreEV} are benchmarks to evaluate video captioning metrics. VATEX-EVAL measures the correlation between metric scores and human rankings of captions, while ActivityNet-FOIL tests whether a metric can distinguish between correct captions and those containing hallucinations, artificially generated by humans.





\begin{figure*}[t]
 \centering
 \includegraphics[width=\textwidth]{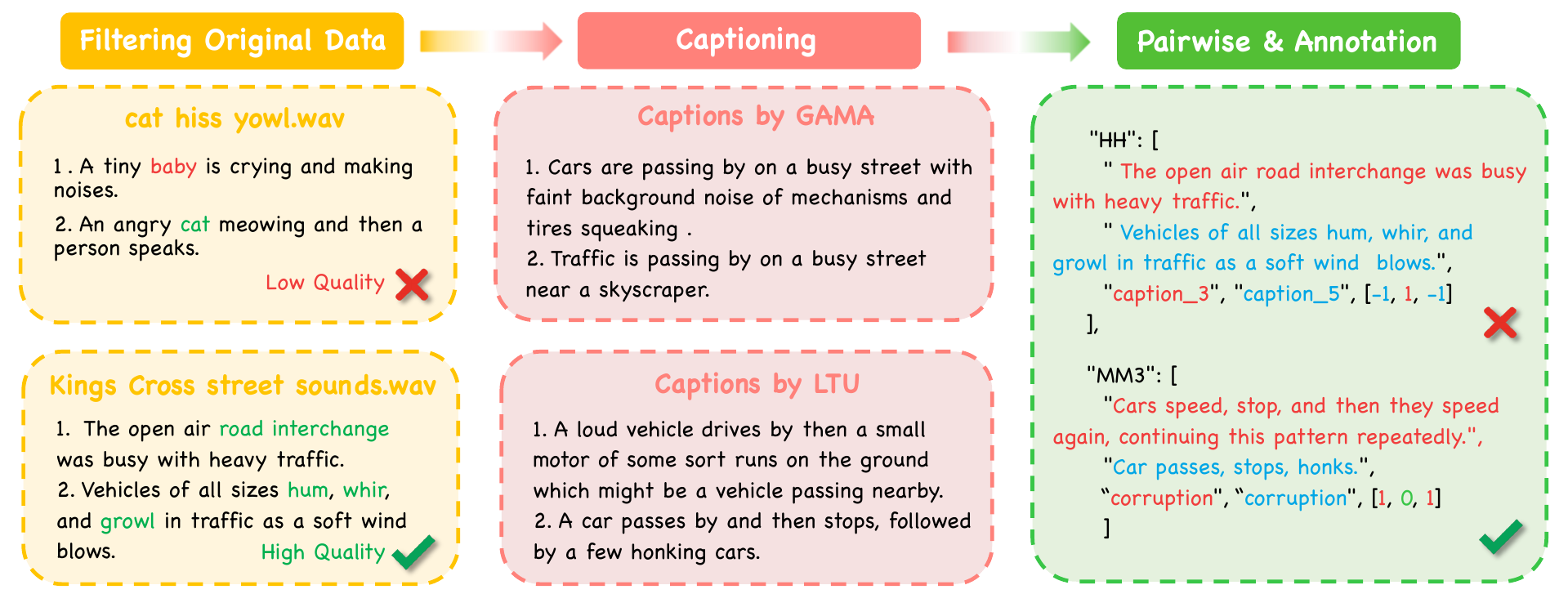}
 \caption{The process of constructing the BRACE-Main benchmark begins with filtering high-quality audio-caption pairs for further processing. Next, we utilize LALMs to generate captions, subsequently corrupting a portion of them. Then, we pair the audio-caption data and have human annotators manually annotate the pairs. Finally, we select the data for which the annotators reach a consensus.}
 \label{fig: Main-Process}
 \vspace{-4mm}
\end{figure*}




\section{BRACE Dataset Construction}
To construct a high-quality benchmark for evaluating reference-free ACEM, we developed two benchmarks: \textbf{BRACE-Main} is developed to evaluate how well CLAPScore correlates with human judgments when assessing the quality of diverse types of captions. \textbf{BRACE-Hallucination} is designed to measure CLAPScore’s sensitivity to hallucinated content within captions. Both benchmarks can also be extended to assess the audio-caption alignment capabilities of LALMs.

\subsection{BRACE-Main Benchmark} \label{sec:brace-main benchmark}

\subsubsection{BRACE-Main Construction}
To construct a challenging audio pairwise comparison benchmark, we first conduct source data selection and filtering to obtain high-quality dataset.

\paragraph{Source Data Selection and Filtering} \label{sec:source data selection and filtering}We selected the commonly used AudioCaps~\cite{kim2019audiocaps} and Clotho~\cite{drossos2020clotho} evaluation datasets as our source data for audio captioning. As shown in Figure~\ref{fig: Main-Process}, we observed that some audio clips had captions of lower quality, making it difficult to determine which ones accurately described the audio content. To address this, we used Qwen2.5-7B-Instruct~\cite{qwen2.5} to filter out audio clips with excessive semantic variation between captions, ensuring higher consistency. This filtering process also reduces the need for extensive human annotation, thereby lowering costs and improving the reliability of the data. The filtering prompt is provided in Figure~\ref{fig: data_filter_prompt}. Ultimately, we retained 765 audio clips from AudioCaps evaluation dataset and 1262 audio clips from Clotho evaluation dataset. 

After filtering out the low quality audio-caption pairs, we construct the high-quality pairwise audio captions using the following technique.

\paragraph{Dataset Construction} We aim to comprehensively evaluate CLAP’s ability to assess caption quality and the alignment capability of LALM for audio captions, focusing on semantic alignment and grammatical correctness. To this end, we construct three types of audio-caption pairs, as shown in Table~\ref{table: six pairs and explanation}, where HH stands for Human-Human comparison, HM stands for Human-Machine comparison, MM stands for Machine-Machine comparison. 
\textbf{Human} is obtained from well-filtered captions from AudioCaps and Clotho. \textbf{Generated} is obtained by using LALMs such as LTU~\cite{Gong2023ListenTA} and GAMA~\cite{Ghosh2024GAMAAL}. \textbf{Corruption} is derived using large language models to create low-quality text. Specifically, we use Qwen2.5-7B-Instruct to shorten the captions to fewer than five words. This approach has two main advantages. First, shorter captions are less likely to capture the full semantic meaning of the original audio, creating semantically corrupted data. Second, we found that Qwen2.5-7B-Instruct performs poorly when identifying sentence components in longer captions. Therefore, we shorten the captions to facilitate the introduction of fluency errors\cite{Zhou2021CanAC} such as incomplete sentences. The model intentionally introduces these errors during caption corruption. This corruption process ensures the creation of low-quality captions that exhibit both semantic and grammatical flaws. Consequently, it enables us to rigorously evaluate CLAP models' sensitivity to caption quality in both aspects. The prompt we use for data corruption is as shown in Figure~\ref{fig: data_corruption_prompt}.

After the construction of the audio-caption pairs, we find experienced human annotators to further annotate our data.
\paragraph{Data Annotation} To achieve good performance in our benchmark, for each clip, three annotators chose the caption that best aligns with the audio, if the first is better, the annotator will score 1, otherwise -1. If both captions were deemed equally appropriate, annotators marked 0.  Finally, we summed the three individual scores to obtain the total score, which serves as the human annotation for the data. To ensure the quality of our benchmarks, we select annotators from one of the top universities in China. 

After the data is annotated, we conduct a further data filtering step for high-quality data in which the three annotators reach a consensus. 

\paragraph{Further Filtering}
The total score of each caption pair ranges from -3 to 3.
As illustrated in Figure~\ref{fig: Distribution of Scores by Dataset}, 54.1\% of caption pairs in AudioCaps and 62.6\% in Clotho have absolute human scores of 2 or higher, indicating that at least two annotators agreed on which caption better aligns with the audio. Filtering out pairs with lower scores improves consistency. As shown in Table~\ref{Fleiss-Kappa-table}, the Fleiss-Kappa score significantly improves, with AudioCaps increasing from 0.38 to 0.98 and Clotho increasing from 0.44 to 0.84. This demonstrates better inter-annotator agreement after filtering, further confirming the high quality of our benchmark.

\begin{figure}[htbp]
  \vspace{-3mm}
  \centering
  \begin{minipage}[t]{0.50\textwidth}
    \centering
    \includegraphics[width=\textwidth]{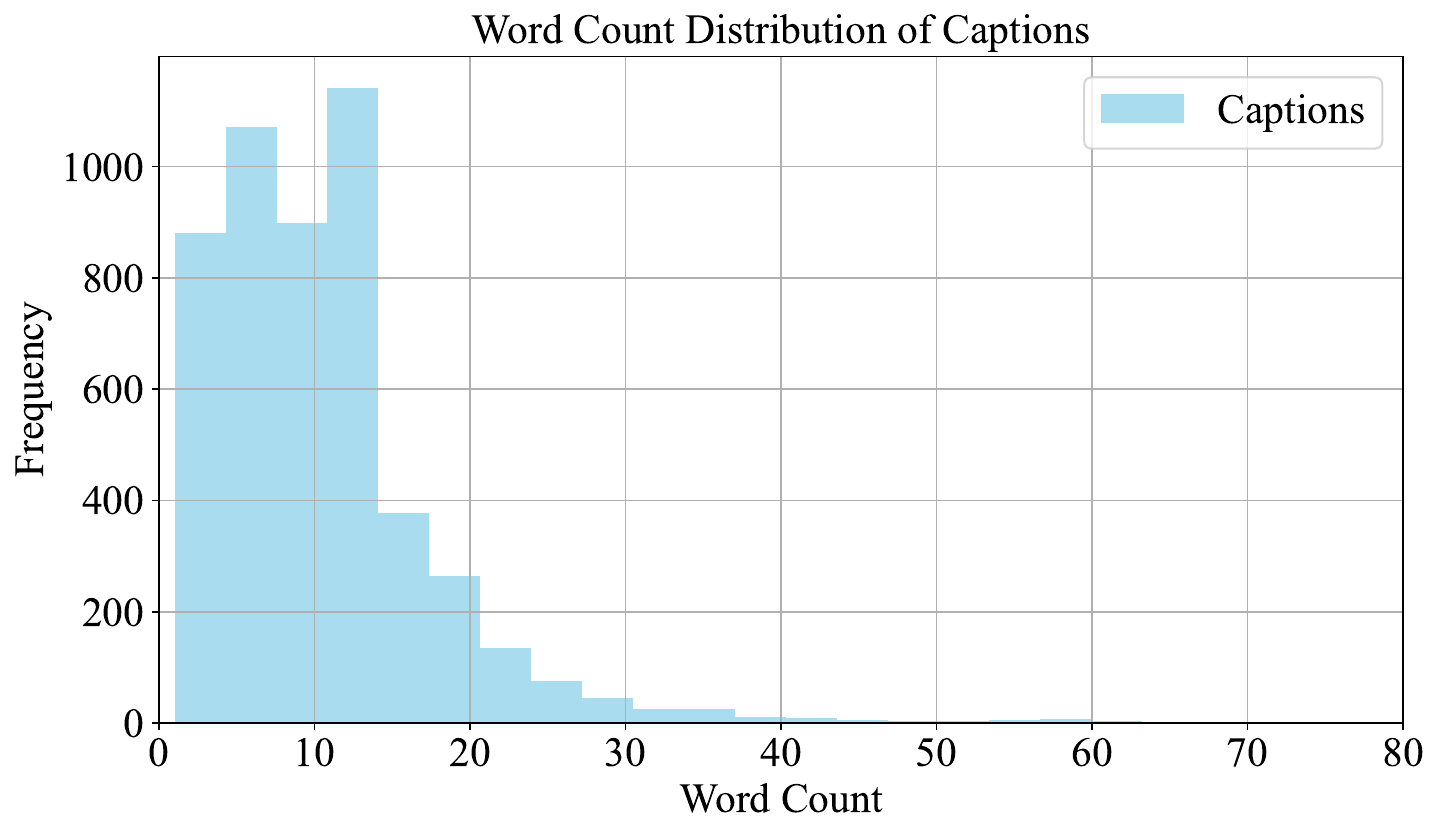}
    \caption{Word count distribution of BRACE-Main. Most captions exhibit a word count concentrated within 20 words.}
    \label{fig:word_count_distribution_main}
  \end{minipage}
  \hfill
  \begin{minipage}[t]{0.48\textwidth}
    \centering
    \includegraphics[width=0.6\textwidth]{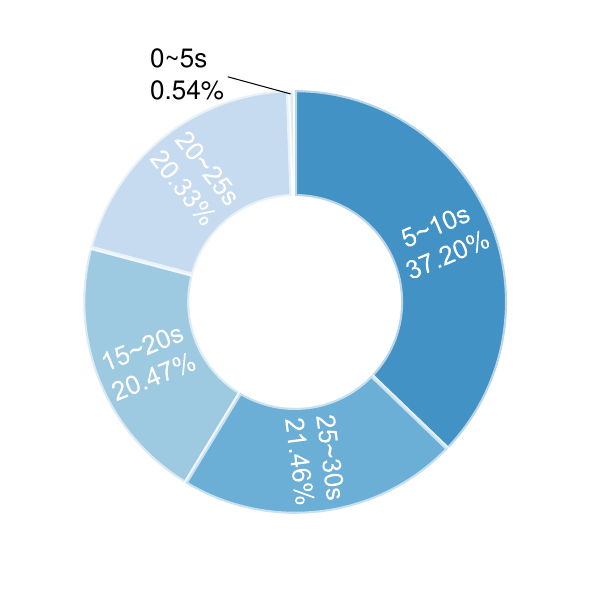}
    \caption{Audio duration distribution of BRACE. Most of the audio samples falling within the 5 to 30 seconds range.}
    \label{fig:audio_duration_distribution}
  \end{minipage}
  \vspace{-4mm}
\end{figure}

\subsubsection{BRACE-Main Data Overview}
After filtering, the total number of retained caption pairs was 2496. As shown in Figure~\ref{fig:word_count_distribution_main}, most captions have a word count concentrated within 20 words, indicating a high degree of consistency in the length of the captions across the dataset. Furthermore, Figure~\ref{fig:audio_duration_distribution} reveals that BRACE-Main contains a rich variety of audio lengths, with most of the audio samples falling within the 5 to 30 seconds range. This suggests that our benchmark encompasses a wide variety of audio lengths, ensuring the completeness and reliability of the evaluation.

\subsection{BRACE-Hallucination Benchmark} \label{sec:brace-hallucination benchmark}
\paragraph{Dataset Construction} We utilized the filtered audio clips, as outlined in Section~\ref{sec:source data selection and filtering}, totaling 2027 clips for our BRACE-Hallucination track. 
We leverage a large language model in a few-shot setting to randomly select and replace a noun within a sentence. The replacement noun is chosen according to two key criteria: 
First, it must fit naturally within the sentence, ensuring the overall sentence remains coherent and logically sound. Second, it must differ significantly in meaning from the original noun, introducing a clear change in the sentence's context.

Our prompt includes not only illustrative examples, but also detailed explanations of the underlying substitution principles demonstrated by each example.
In contrast to the BRACE-Main track, which uses Qwen2.5-7B-Instruct, we employ the GPT-4o \cite{openai2023gpt} model for hallucinated data generation. Since the GPT-4o model handles longer prompts more effectively, ensuring that the generated hallucination align with our requirements. In total, we obtained 10315 caption pairs. The prompt used is shown in Figure~\ref{fig: hallucination_generator_prompt}.

\paragraph{Data Analysis} 
To further analyze the statistics of BRACE-Hallucination, we examined the caption lengths and audio statistics. The average caption length in BRACE-Hallucination is 10.78. Furthermore, 94.93\% of the captions and their corresponding hallucinated captions have identical lengths. This indicates that the model's performance on BRACE-Hallucination is not strongly influenced by caption length. Additionally, the distribution of audio lengths is consistent with that of BRACE-Main, as depicted in Figure~\ref{fig:audio_duration_distribution}.

\begin{table}[t]
  \caption{Performance of CLAPs, SLIDE-CLAPs and LALMs on BRACE-Main. Results are shown across different caption pair types and overall. The best-performing models in each category are highlighted in \textbf{bold}, and the second-best scores are \underline{underlined}. For CLAP we present the average performance across 10 independent runs.}
  \label{tab:Main-Results-on-BRACE-Main}
  \centering
  \setlength{\tabcolsep}{3.5pt}
  \begin{tabular}{lccccccccc}
    \toprule
    & \multicolumn{4}{c}{\textbf{AudioCaps}} & \multicolumn{4}{c}{\textbf{Clotho}} & \textbf{Avg-All} \\
    \cmidrule(lr){2-5} \cmidrule(lr){6-9}
    \textbf{Model} & HH & HM & MM & All & HH & HM & MM & All & \\
    \midrule
    \multicolumn{10}{c}{\textbf{CLAP}} \\
    \midrule
    M2D-CLAP      & 47.96            & \underline{70.18}& \underline{60.41} & \underline{62.96}   & 49.24            & 56.11             & 58.66               & 56.61          & 59.78\\
    MS-CLAP-2022  & 57.75            & 48.84            & 59.45             & 54.93               & 46.96            & \textbf{85.03}    & \underline{62.30}   & \textbf{69.13} & \underline{62.03}\\
    MS-CLAP-2023  & \textbf{61.75}   & 52.71            & 52.33             & 53.56               & \textbf{57.30}   & \underline{74.73} & \textbf{64.26}      & \underline{67.58} & 60.57\\
    LAION-CLAP    & \underline{60.63}& \textbf{85.87}   & \textbf{65.35}    & \textbf{73.33}      & \underline{56.29}& 70.13             & 62.03               & 64.54          & \textbf{68.93} \\
    \midrule
    \multicolumn{10}{c}{\textbf{SLIDE-CLAP}} \\
    \midrule
    M2D-CLAP          & 47.76 & \underline{71.55} & \underline{61.60} & \underline{64.03} & 50.70 & 57.66 & 59.47 & 57.79 & 60.91 \\
    MS-CLAP-2022      & \underline{60.47} & 48.03 & 59.77 & 55.05 & 48.37 & \textbf{87.45} & 63.00 & \textbf{70.56} & \underline{62.81} \\
    MS-CLAP-2023      & \textbf{66.12} & 52.63 & 52.47 & 54.06 & \textbf{60.29} & \underline{76.89} & \textbf{64.59} & \underline{68.96} & 61.51 \\
    LAION-CLAP        & 59.84 & \textbf{86.08} & \textbf{66.92} & \textbf{74.13} & \underline{55.32} & 71.63 & \underline{63.76} & 65.89 & \textbf{70.01} \\
    \midrule
    \multicolumn{10}{c}{\textbf{LALM}} \\
    \midrule
    AF2            & \textbf{65.26} & \underline{68.97} & \textbf{60.99} & \textbf{64.70} & \textbf{56.11} & \underline{63.30} & \textbf{61.83} & \textbf{61.68} & \textbf{63.19} \\
    LTU               & \underline{60.67} & 63.41  & \underline{59.97} & \underline{61.44}  & 51.31  & 59.12  & 57.76  & 57.54  & 59.49 \\
    GAMA              & 0.00  & 16.47  & 8.60  & 11.04  & 13.48  & 16.00  & 12.90  & 14.19  & 12.62 \\

    Qwen-Audio-Chat   & 49.61 & 62.18  & 59.21 & 59.42  & \underline{55.21}  & \textbf{65.49} & \underline{59.15} & \underline{61.10} & \underline{60.26} \\
    Qwen2-Audio-Instruct & 52.38 & 55.25 & 48.39 & 51.79 & 48.05 & 55.78 & 55.75 & 54.83 & 53.31 \\
    GPT-4o-Audio-Preview & 60.22 & \textbf{71.38} & 51.96 & 58.33 & 50.62 & 59.11 & 49.43 & 52.14 & 55.24 \\
    \bottomrule
  \end{tabular}
  \vspace{-4mm}
\end{table}

\section{Experiments}
This chapter presents the experimental study in detail. Section \ref{sec: experiment_setting} outlines the model configurations and evaluation metrics. Section \ref{sec: main_results} reports the experimental results. We then conduct a detailed analysis of the results, analyzing the limitations of the CLAP and LALM models separately in Section \ref{sec: limitations of claps} and Section \ref{sec: limitations of lalms}, where we present several examples and discuss the constraints of their current capabilities. Notably, our benchmark provides a comprehensive and challenging evaluation setting that effectively reveals nuanced weaknesses in existing models, offering valuable insights for future improvements.

\subsection{Experimental settings} \label{sec: experiment_setting}

\paragraph{Metric} For BRACE evaluation, we use strategy-specific methods. CLAP computes similarity between audio and captions, while SLIDE-CLAP enhances its stability via sliding window averaging. LALM evaluates caption pairs through prompt-based preference selection, utilizing diverse prompt templates and a secondary model for final choices. 
For LALMs, we design three prompting levels: \textit{naive}, \textit{simple}, and \textit{complex}, with increasing complexity. Each level includes \textit{tie} and \textit{non-tie} variants, indicating whether the tie option is provided to the model.
We report results on BRACE-Main and BRACE-Hallucination benchmarks.
Full details about models' evaluation strategy are provided in Appendix~\ref{Appendix: brace evaluation strategies}.

\paragraph{CLAP} We evaluate several mainstream CLAP models, including MS-CLAP-2022~\cite{elizalde2023clap}, MS-CLAP-2023~\cite{elizalde2024natural}, M2D-CLAP~\cite{niizumi2024m2d-clap}, and LAION-CLAP~\cite{wu2024largescalecontrastivelanguageaudiopretraining}, using their best-performing configurations.

\paragraph{SLIDE-CLAP} SLIDE-CLAP utilizes the same base CLAP models but incorporates a sliding window technique for improved stability. The window size is determined by the fixed input length required by each audio encoder: 5 seconds for MS-CLAP-2022, 7 seconds for MS-CLAP-2023, and 10 seconds for both M2D-CLAP and LAION-CLAP. A uniform hop size of 1 second is applied across all models.

\paragraph{LALM} We evaluate the following LALM models: LTU~\cite{Gong2023ListenTA}, GAMA~\cite{Ghosh2024GAMAAL}, Qwen-Audio-Chat~\cite{Qwen-Audio}, AF2 (Audio Flamingo 2)~\cite{ghosh2025audioflamingo2audiolanguage}, Qwen2-Audio-Instruct~\cite{Qwen2-Audio} and GPT-4o-Audio-Preview\cite{achiam2023gpt}, all using default settings. To ensure determinism and reproducibility, the generation temperature is fixed at 0.

\begin{figure}[t]
  \centering
  \begin{minipage}[t]{0.45\textwidth}
    \captionof{table}{Performance of CLAPs, SLIDE-CLAPs and LALMs on BRACE-Hallucination. The best-performing models in each category are highlighted in \textbf{bold}, and the second-best scores are \underline{underlined}. Both variants of CLAP demonstrate superior performance compared to LALM.}
    \vspace{-2.3mm}
    \label{tab:Main-Results-on-BRACE-Hallu}
    \resizebox{\textwidth}{!}{
      \begin{tabular}{lccc}
        \toprule
        \textbf{Model} & \textbf{AudioCaps} & \textbf{Clotho} & \textbf{Avg-All} \\
        \midrule
        \multicolumn{4}{c}{\textbf{CLAP}} \\
        \midrule
        M2D-CLAP       & \textbf{90.47}     & 81.91             & \textbf{86.19} \\
        MS-CLAP-2022   & 74.43              & \textbf{88.66}    & 81.55 \\
        MS-CLAP-2023   & 79.15              & \underline{83.45} & 81.30 \\
        LAION-CLAP     & \underline{86.99}  & 78.88             & \underline{82.94} \\
        \midrule
        \multicolumn{4}{c}{\textbf{SLIDE-CLAP}} \\
        \midrule
        M2D-CLAP             & \textbf{91.50} & 85.02 & \textbf{88.26} \\
        MS-CLAP-2022         & 78.86 & \textbf{93.46} & \underline{86.16} \\
        MS-CLAP-2023         & 84.12 & \underline{87.85} & 85.99 \\
        LAION-CLAP           & \underline{87.79} & 80.95 & 84.37 \\
        \midrule
        \multicolumn{4}{c}{\textbf{LALM}} \\
        \midrule
        AF2               & 79.55 & 72.91 & 76.23 \\
        LTU                  & 63.35 & 59.63 & 61.49 \\
        GAMA                 & 18.22 & 19.35 & 18.79 \\
        Qwen-Audio-Chat      & \underline{79.85} & \underline{74.64} & \underline{77.25} \\
        Qwen2-Audio-Instruct & 61.17 & 57.76 & 59.47 \\
        GPT-4o-Audio-Preview & \textbf{95.76} & \textbf{96.75} & \textbf{96.37} \\
        \bottomrule
      \end{tabular}
    }
  \end{minipage}
  \hfill
  \begin{minipage}[t]{0.514\textwidth}
    \label{fig:examples_1}
    \resizebox{\textwidth}{!}{
        \centering
        \includegraphics[width=\textwidth]{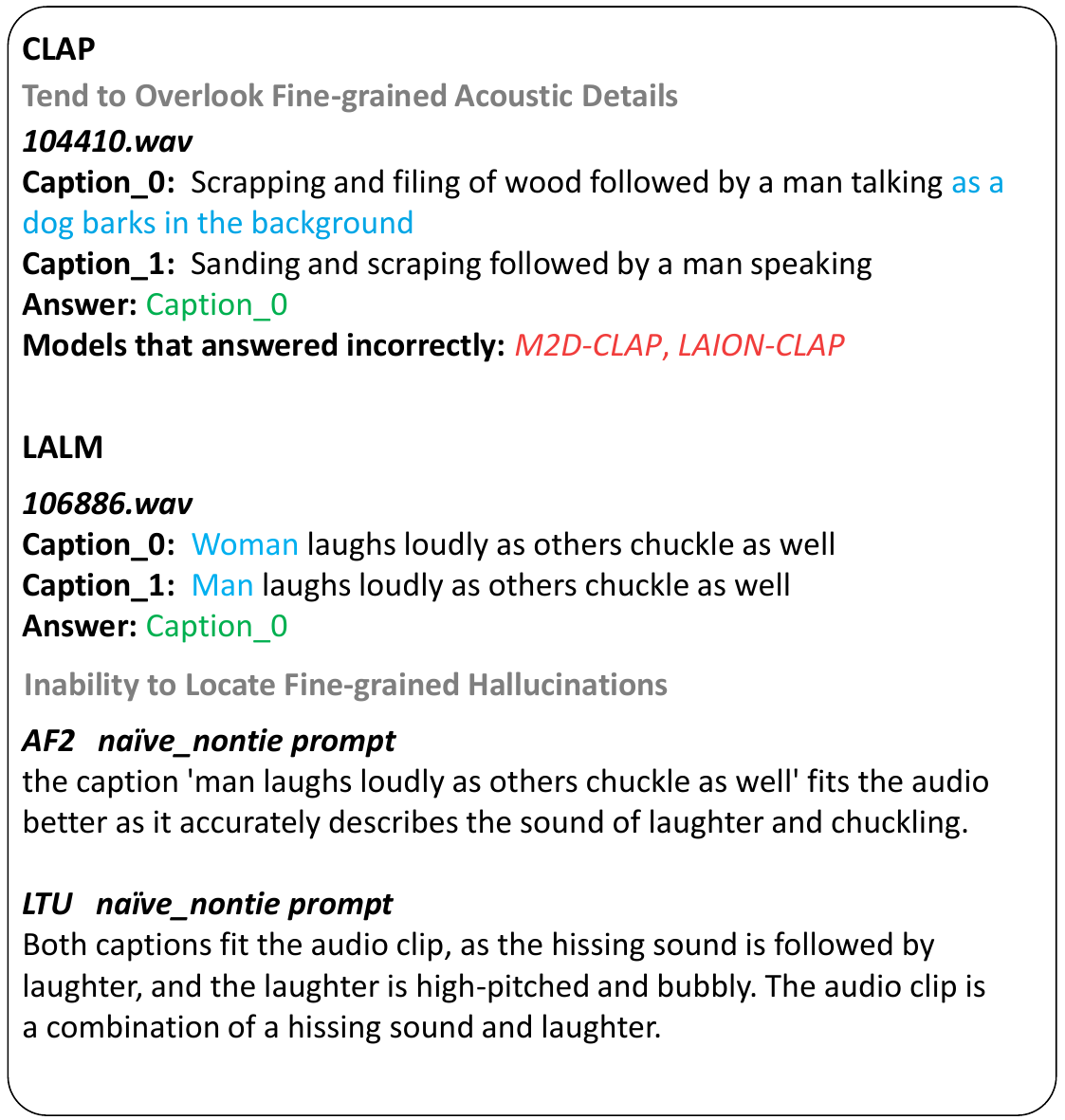}
    }
    \caption{Representative examples from model evaluation}
    \label{fig:examples}
    \end{minipage}
    \vspace{-4mm}
\end{figure}

\subsection{Main results} \label{sec: main_results}
Table \ref{tab:Main-Results-on-BRACE-Main} and Table \ref{tab:Main-Results-on-BRACE-Hallu} compare the results of various LALMs and CLAPs on the BRACE benchmark. Our key findings are:

\paragraph{The benchmark poses a significant challenge and supports effective meta-evaluation.} On BRACE-Main, the best-performing model LAION-CLAP achieves an F1-score of 70.01, while others range from $\sim$55 to 70 depending on architecture and subset. On BRACE-Hallucination, the top-performing model M2D-CLAP reaches an F1-score of 88.26, though performance still varies significantly across models. These results show that our benchmark effectively differentiates model quality and can be used as a reliable meta-evaluation tool to select robust CLAP-based metrics. 

\paragraph{The limited input window size of CLAP leads to performance instability.} As shown in Table \ref{tab:Performance-of-CLAPs-Updated}, the constrained input window length in CLAP models contributes to inconsistent performance, particularly when processing longer audio clips. Since CLAP models operate on short fixed-length segments, longer audio inputs must be truncated or sampled. This stochastic truncation introduces inconsistency across runs, undermining the stability and reproducibility of evaluation results. To address this issue, we introduce a sliding window strategy, where the CLAP embeddings from each audio segment are averaged across overlapping windows. This approach improves the stability of the model's output, as shown in Figure~\ref{fig: slide}. 

\begin{figure}[t]
  \centering
  \includegraphics[width=\textwidth]{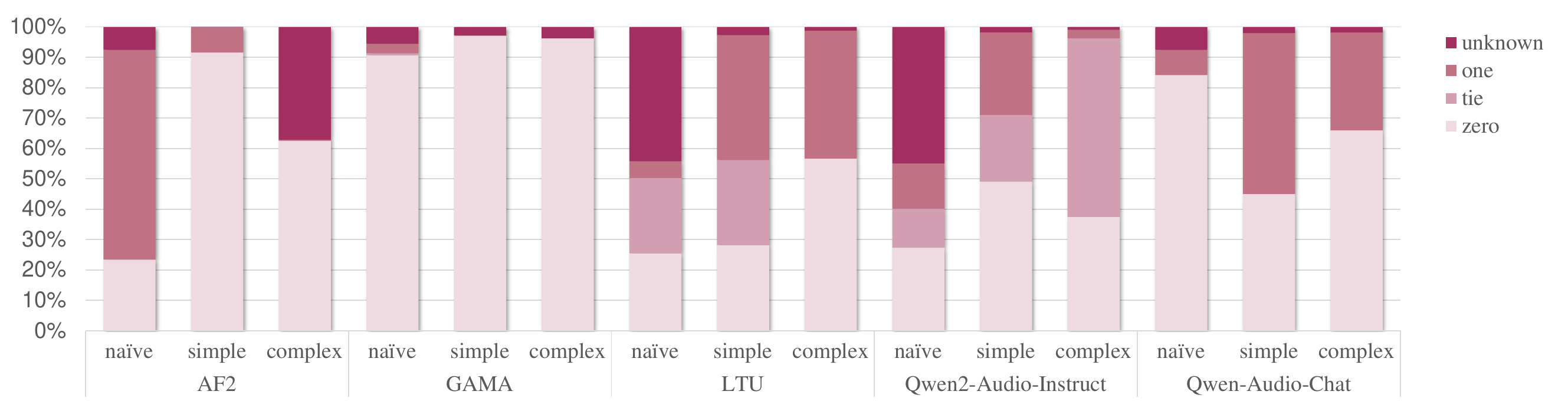}
  \caption{Output distribution of open-source LALMs on BRACE-Hallucination using prompts of different complexity levels (\textit{naive}, \textit{simple}, \textit{complex}). The labels "zero", "one", "tie", "unknown" represent the proportions of the model’s outputs choosing \texttt{caption\_0}, \texttt{caption\_1}, ties or invalid responses, respectively. Notably, GAMA shows a clear imbalance, strongly favoring \texttt{caption\_0}, which clearly reflects a position bias problem.}
  \label{fig:lalm-hallu-distribution}
  \vspace{-6mm}
\end{figure}

\paragraph{CLAP-Based Metrics Show Inconsistent Performance across Different Caption Pair Types} Compared to other types, SLIDE-CLAP models perform best on Human-Machine (HM) pairs, likely because stylistic differences between captions make it easier to determine which is superior. 
However, all CLAP models perform poorly on Human-Human (HH) and Machine-Machine (MM) comparisons, with none achieving a score above 70 on either type. This indicates that CLAP struggles to identify fine-grained distinctions between high-quality human-written captions or between captions produced by similar models, suggesting substantial room for improvement. 


\paragraph{Performance Disparities Between Open- and Closed-Source LALMs on BRACE Benchmarks.} On the BRACE-Main benchmark, both open- and closed-source LALMs underperform compared to CLAP-based models. The strongest open-source LALM, AF2, achieves an F1-score of 63.19, while the closed-source GPT-4o-Audio-Preview reaches 55.24. In contrast, several other models (e.g., GAMA) perform poorly, with scores dropping to 20 or below on certain subsets. On the BRACE-Hallucination benchmark, however, the performance gap between closed- and open-source models becomes substantially more pronounced. GPT-4o-Audio-Preview attains a state-of-the-art F1-score of 96.37, whereas the second-best, open-source Qwen-Audio-Chat, reaches only 77.25. These results highlight the significant disparity in fine-grained hallucination detection between closed- and open-source LALMs.

Since SLIDE-CLAP aggregates more comprehensive audio information through a sliding window, it achieves better performance and greater stability compared to the standard CLAP on the BRACE benchmark. Because of these improvements, we refer to the sliding window enhanced version (SLIDE-CLAP) simply as CLAP throughout the rest of this paper for brevity.

\subsection{Insufficient Acoustic Granularity and Comprehensiveness of CLAP Models} \label{sec: limitations of claps}
In this section, we present our analysis of the limitations of CLAP on the BRACE benchmark. Based on this comprehensive evaluation, we have drawn the following conclusions. 

\paragraph{Tend to Overlook Fine-grained Acoustic Details} As shown in Figure~\ref{fig:examples}, CLAP models tend to capture coarse-grained semantic information within the input audio window, often focusing on dominant acoustic events or salient foreground sounds. However, it frequently overlooks fine-grained acoustic details, such as subtle background cues.


\paragraph{Syntactic Oversight in CLAP-based Retrieval} CLAP models predominantly align audio and text at the semantic level, focusing on the presence of acoustically salient words such as sound sources or events. However, it often overlooks syntactic structure and fluency errors, such as incomplete sentences, missing conjunctions and so on. As a result, captions that are semantically relevant but syntactically incorrect or fragmented may receive higher similarity scores. This indicates that CLAP lacks sensitivity to the grammatical well-formedness of captions, which is essential for capturing coherent and contextually faithful descriptions of audio events. More examples about CLAP models' failure cases can be seen in Appendix~\ref{case-study}.

Future work may improve CLAP models by integrating fine-grained acoustic features and syntax-aware training to enhance grammatical alignment.
A syntax- and acoustics-sensitive CLAP model can support reference-free evaluation by filtering semantically flawed captions, thereby improving dataset quality.

\begin{table}
\centering
\caption{Performance of LALMs on BRACE-Main and BRACE-Hallucination. Prompt configurations vary along two dimensions: complexity (\textit{naive}, \textit{simple}, \textit{complex}) and tie setting (\textit{non-tie}, \textit{tie}).}
\label{tab:lalm-main-and-hallu}
\begin{tabular}{l*{6}{c}}
\toprule
& \multicolumn{3}{c}{\textbf{Non-Tie}} & \multicolumn{3}{c}{\textbf{Tie}} \\
\cmidrule(lr){2-4} \cmidrule(lr){5-7}
\textbf{Model} & \textbf{Naive} & \textbf{Simple} & \textbf{Complex} & \textbf{Naive} & \textbf{Simple} & \textbf{Complex} \\
\midrule
\multicolumn{7}{c}{\textbf{BRACE-Main}} \\
\midrule
AF2 & 62.94 & 20.39 & 1.35 & \textbf{63.19} & 22.03 & 3.26 \\
LTU & 16.55 & \textbf{58.82} & 8.09 & 12.15 & 43.02 & 46.62 \\
GAMA & \textbf{10.69} & 9.81 & 0.67 & 3.97 & 2.26 & 0.00 \\
Qwen-Audio-Chat & 28.99 & \textbf{59.64} & 50.74 & 17.69 & 59.84 & 49.42 \\
Qwen2-Audio-Instruct & 37.75 & 51.12 & 48.03 & 20.73 & \textbf{52.73} & 17.74 \\
GPT-4o-Audio-Preview & 68.30 & 26.35 & 23.56 & \textbf{69.50} & 36.18 & 30.52 \\
\midrule
\multicolumn{7}{c}{\textbf{BRACE-Hallucination}} \\
\midrule
AF2 & \textbf{76.23} & 32.32 & 0.71 & 72.34 & 27.88 & 1.18 \\
LTU & 11.43 & \textbf{61.50} & 8.25 & 7.75 & 41.45 & 48.96 \\
GAMA & \textbf{18.79} & 1.19 & 0.61 & 9.33 & 0.45 & 0.03 \\
Qwen-Audio-Chat & 36.57 & \textbf{77.00} & 63.05 & 24.66 & 76.72 & 64.10 \\
Qwen2-Audio-Instruct & 45.61 & \textbf{59.46} & 53.15 & 28.56 & 51.67 & 6.76 \\
GPT-4o-Audio-Preview & 98.29 & 86.18 & 79.75 & \textbf{98.59} & 91.77 & 88.62 \\
\bottomrule
\end{tabular}
\vspace{-6mm}
\end{table}

\subsection{LALMs Suffer from Comparing Audio Caption Quality} \label{sec: limitations of lalms}
In this section, we systematically analyze the performance of LALM models on the BRACE benchmark and identify several core limitations that inform future research directions.

\paragraph{Poor Instruction Understanding and Following} LALMs exhibit a noticeable decline in performance as prompt complexity increases, even when provided with more detailed comparison criteria and clearer problem definitions. Simpler prompts tend to yield better results, as shown in Table~\ref{tab:lalm-main-and-hallu}. Additionally, even when the output is structured through multiple-choice formats, models still make incorrect choices, such as outputting "none" inappropriately.

\paragraph{Position Bias in Multiple Prompt Templates} As shown in Figure~\ref{fig:lalm-hallu-distribution}, models like AF2 or GAMA often activate fixed patterns from training when using various prompt templates, leading to a position bias in their outputs. Specifically, these models select \texttt{caption\_0} or \texttt{caption\_1} without considering the actual content of the caption or audio. This pattern reflects a lack of genuine understanding and reasoning, relying instead on positional cues. 

\paragraph{Inability to Locate Fine-grained Hallucinations} In our analysis of LALMs' errors on the BRACE-Hallucination task, some models attempted to provide explanations for their selections but failed to accurately locate the hallucinated content within the given input. Several examples are provided in Appendix~\ref{case-study}, highlighting these errors.

 To address these issues, key directions include strengthening instruction following and reasoning under complex prompts, applying debiasing techniques to reduce positional bias, and improving fine-grained hallucination detection by enabling more comprehensive recognition of sound events in challenging input scenarios.

\section{Conclusion}
We introduce BRACE, a benchmark designed for the systematic evaluation of reference-free ACEMs and LALMs. BRACE consists of two sub-benchmarks: BRACE-Main and BRACE-Hallucination, which are constructed through a combination of LLM-based generation, corruption, and expert annotation. BRACE measures the quality of metrics by assessing the alignment between reference-free ACEMs and human judgments. It also highlights inherent issues with fine-grained perception and limited sensitivity to syntax and grammar in CLAP-based metrics. In contrast, testing LALMs exposes their difficulties with poor instruction understanding, position bias, and similar issues, offering valuable diagnostic insights for the future development of LALMs. We aim for BRACE to drive progress in audio-language evaluation and model development, leading to more accurate and robust metrics and models.

\section{Acknowledgments}
This work is supported by the National Key R\&D Program of China (2024YFA1014003), National Natural Science Foundation of China (92470121, 62402016), and High-performance Computing Platform of Peking University.
{\small
\bibliographystyle{ieeetr}
\bibliography{main}

@inproceedings{Papineni2002BleuAM,
  title={Bleu: a Method for Automatic Evaluation of Machine Translation},
  author={Kishore Papineni and Salim Roukos and Todd Ward and Wei-Jing Zhu},
  booktitle={Annual Meeting of the Association for Computational Linguistics},
  year={2002},
  url={https://api.semanticscholar.org/CorpusID:11080756}
}

@inproceedings{Lin2004ROUGEAP,
  title={ROUGE: A Package for Automatic Evaluation of Summaries},
  author={Chin-Yew Lin},
  booktitle={Annual Meeting of the Association for Computational Linguistics},
  year={2004},
  url={https://api.semanticscholar.org/CorpusID:964287}
}

@inproceedings{Banerjee2005METEORAA,
  title={METEOR: An Automatic Metric for MT Evaluation with Improved Correlation with Human Judgments},
  author={Satanjeev Banerjee and Alon Lavie},
  booktitle={IEEvaluation@ACL},
  year={2005},
  url={https://api.semanticscholar.org/CorpusID:7164502}
}

@article{Vedantam2014CIDErCI,
  title={CIDEr: Consensus-based image description evaluation},
  author={Ramakrishna Vedantam and C. Lawrence Zitnick and Devi Parikh},
  journal={2015 IEEE Conference on Computer Vision and Pattern Recognition (CVPR)},
  year={2014},
  pages={4566-4575},
  url={https://api.semanticscholar.org/CorpusID:9026666}
}

@article{Anderson2016SPICESP,
  title={SPICE: Semantic Propositional Image Caption Evaluation},
  author={Peter Anderson and Basura Fernando and Mark Johnson and Stephen Gould},
  journal={ArXiv},
  year={2016},
  volume={abs/1607.08822},
  url={https://api.semanticscholar.org/CorpusID:11933981}
}

@article{Liu2016ImprovedIC,
  title={Improved Image Captioning via Policy Gradient optimization of SPIDEr},
  author={Siqi Liu and Zhenhai Zhu and Ning Ye and Sergio Guadarrama and Kevin P. Murphy},
  journal={2017 IEEE International Conference on Computer Vision (ICCV)},
  year={2016},
  pages={873-881},
  url={https://api.semanticscholar.org/CorpusID:3873857}
}

@article{Bhosale2022TexttoAudioGB,
  title={Text-to-Audio Grounding Based Novel Metric for Evaluating Audio Caption Similarity},
  author={Swapnil Bhosale and Rupayan Chakraborty and Sunil Kumar Kopparapu},
  journal={ArXiv},
  year={2022},
  volume={abs/2210.06354},
  url={https://api.semanticscholar.org/CorpusID:252846331}
}

@article{Zhou2021CanAC,
  title={Can Audio Captions Be Evaluated With Image Caption Metrics?},
  author={Zelin Zhou and Zhiling Zhang and Xuenan Xu and Zeyu Xie and Mengyue Wu and Kenny Q. Zhu},
  journal={ICASSP 2022 - 2022 IEEE International Conference on Acoustics, Speech and Signal Processing (ICASSP)},
  year={2021},
  pages={981-985},
  url={https://api.semanticscholar.org/CorpusID:238582669}
}

@article{Zhang2019BERTScoreET,
  title={BERTScore: Evaluating Text Generation with BERT},
  author={Tianyi Zhang and Varsha Kishore and Felix Wu and Kilian Q. Weinberger and Yoav Artzi},
  journal={ArXiv},
  year={2019},
  volume={abs/1904.09675},
  url={https://api.semanticscholar.org/CorpusID:127986044}
}

@article{Wijngaard2023ACESEA,
  title={ACES: Evaluating Automated Audio Captioning Models on the Semantics of Sounds},
  author={Gijs Wijngaard and Elia Formisano and Bruno L. Giordano and Michel Dumontier},
  journal={2023 31st European Signal Processing Conference (EUSIPCO)},
  year={2023},
  pages={770-774},
  url={https://api.semanticscholar.org/CorpusID:268724100}
}

@article{Elizalde2023CLAPLA,
  title={CLAP Learning Audio Concepts from Natural Language Supervision},
  author={Benjamin Elizalde and Soham Deshmukh and Mahmoud Al Ismail and Huaming Wang},
  journal={ICASSP 2023 - 2023 IEEE International Conference on Acoustics, Speech and Signal Processing (ICASSP)},
  year={2023},
  pages={1-5},
  url={https://api.semanticscholar.org/CorpusID:249605738}
}

@article{Shi2021EMScoreEV,
  title={EMScore: Evaluating Video Captioning via Coarse-Grained and Fine-Grained Embedding Matching},
  author={Yaya Shi and Xu Yang and Haiyang Xu and Chunfen Yuan and Bing Li and Weiming Hu and Zhengjun Zha},
  journal={2022 IEEE/CVF Conference on Computer Vision and Pattern Recognition (CVPR)},
  year={2021},
  pages={17908-17917},
  url={https://api.semanticscholar.org/CorpusID:244270365}
}

@article{Gong2023ListenTA,
  title={Listen, Think, and Understand},
  author={Yuan Gong and Hongyin Luo and Alexander H. Liu and Leonid Karlinsky and James Glass},
  journal={ArXiv},
  year={2023},
  volume={abs/2305.10790},
  url={https://api.semanticscholar.org/CorpusID:258762560}
}

@article{Ghosh2024GAMAAL,
  title={GAMA: A Large Audio-Language Model with Advanced Audio Understanding and Complex Reasoning Abilities},
  author={Sreyan Ghosh and Sonal Kumar and Ashish Seth and Chandra Kiran Reddy Evuru and Utkarsh Tyagi and S Sakshi and Oriol Nieto and Ramani Duraiswami and Dinesh Manocha},
  journal={ArXiv},
  year={2024},
  volume={abs/2406.11768},
  url={https://api.semanticscholar.org/CorpusID:270560527}
}

@inproceedings{kim2019audiocaps,
  title={Audiocaps: Generating captions for audios in the wild},
  author={Kim, Chris Dongjoo and Kim, Byeongchang and Lee, Hyunmin and Kim, Gunhee},
  booktitle={Proceedings of the 2019 Conference of the North American Chapter of the Association for Computational Linguistics: Human Language Technologies, Volume 1 (Long and Short Papers)},
  pages={119--132},
  year={2019}
}

@inproceedings{drossos2020clotho,
  title={Clotho: An audio captioning dataset},
  author={Drossos, Konstantinos and Lipping, Samuel and Virtanen, Tuomas},
  booktitle={ICASSP 2020-2020 IEEE International Conference on Acoustics, Speech and Signal Processing (ICASSP)},
  pages={736--740},
  year={2020},
  organization={IEEE}
}

@misc{qwen2.5,
    title = {Qwen2.5: A Party of Foundation Models},
    url = {https://qwenlm.github.io/blog/qwen2.5/},
    author = {{Qwen Team}},
    month = {September},
    year = {2024}
}

@article{achiam2023gpt,
  title={Gpt-4 technical report},
  author={Achiam, Josh and Adler, Steven and Agarwal, Sandhini and Ahmad, Lama and Akkaya, Ilge and Aleman, Florencia Leoni and Almeida, Diogo and Altenschmidt, Janko and Altman, Sam and Anadkat, Shyamal and others},
  journal={arXiv preprint arXiv:2303.08774},
  year={2023}
}

@article{ghosh2023compa,
  title={Compa: Addressing the gap in compositional reasoning in audio-language models},
  author={Ghosh, Sreyan and Seth, Ashish and Kumar, Sonal and Tyagi, Utkarsh and Evuru, Chandra Kiran and Ramaneswaran, S and Sakshi, S and Nieto, Oriol and Duraiswami, Ramani and Manocha, Dinesh},
  journal={arXiv preprint arXiv:2310.08753},
  year={2023}
}

@article{kuan2024understanding,
  title={Understanding Sounds, Missing the Questions: The Challenge of Object Hallucination in Large Audio-Language Models},
  author={Kuan, Chun-Yi and Huang, Wei-Ping and Lee, Hung-yi},
  journal={arXiv preprint arXiv:2406.08402},
  year={2024}
}

@article{kuan2024can,
  title={Can Large Audio-Language Models Truly Hear? Tackling Hallucinations with Multi-Task Assessment and Stepwise Audio Reasoning},
  author={Kuan, Chun-Yi and Lee, Hung-yi},
  journal={arXiv preprint arXiv:2410.16130},
  year={2024}
}

@inproceedings{elizalde2023clap,
  title={Clap learning audio concepts from natural language supervision},
  author={Elizalde, Benjamin and Deshmukh, Soham and Al Ismail, Mahmoud and Wang, Huaming},
  booktitle={ICASSP 2023-2023 IEEE International Conference on Acoustics, Speech and Signal Processing (ICASSP)},
  pages={1--5},
  year={2023},
  organization={IEEE}
}

@inproceedings{elizalde2024natural,
  title={Natural language supervision for general-purpose audio representations},
  author={Elizalde, Benjamin and Deshmukh, Soham and Wang, Huaming},
  booktitle={ICASSP 2024-2024 IEEE International Conference on Acoustics, Speech and Signal Processing (ICASSP)},
  pages={336--340},
  year={2024},
  organization={IEEE}
}

@misc{wu2024largescalecontrastivelanguageaudiopretraining,
      title={Large-scale Contrastive Language-Audio Pretraining with Feature Fusion and Keyword-to-Caption Augmentation}, 
      author={Yusong Wu and Ke Chen and Tianyu Zhang and Yuchen Hui and Marianna Nezhurina and Taylor Berg-Kirkpatrick and Shlomo Dubnov},
      year={2024},
      eprint={2211.06687},
      archivePrefix={arXiv},
      primaryClass={cs.SD},
      url={https://arxiv.org/abs/2211.06687}, 
}

@article{niizumi2024m2d-clap,
    title   = {{M2D-CLAP: Masked Modeling Duo Meets CLAP for Learning General-purpose Audio-Language Representation}},
    author  = {Daisuke Niizumi and Daiki Takeuchi and Yasunori Ohishi and Noboru Harada and Masahiro Yasuda and Shunsuke Tsubaki and Keisuke Imoto},
    journal = {to appear at Interspeech},
    year    = {2024},
    url     = {https://arxiv.org/abs/2406.02032}}

@article{openai2023gpt,
  title={GPT-4 technical report},
  author={OpenAI, R and others},
  journal={ArXiv},
  volume={2303},
  pages={08774},
  year={2023}
}

@inproceedings{panayotov2015librispeech,
  title={Librispeech: an asr corpus based on public domain audio books},
  author={Panayotov, Vassil and Chen, Guoguo and Povey, Daniel and Khudanpur, Sanjeev},
  booktitle={2015 IEEE international conference on acoustics, speech and signal processing (ICASSP)},
  pages={5206--5210},
  year={2015},
  organization={IEEE}
}

@misc{ghosh2025audioflamingo2audiolanguage,
      title={Audio Flamingo 2: An Audio-Language Model with Long-Audio Understanding and Expert Reasoning Abilities}, 
      author={Sreyan Ghosh and Zhifeng Kong and Sonal Kumar and S Sakshi and Jaehyeon Kim and Wei Ping and Rafael Valle and Dinesh Manocha and Bryan Catanzaro},
      year={2025},
      eprint={2503.03983},
      archivePrefix={arXiv},
      primaryClass={cs.SD},
      url={https://arxiv.org/abs/2503.03983}, 
}

@article{Qwen-Audio,
  title={Qwen-Audio: Advancing Universal Audio Understanding via Unified Large-Scale Audio-Language Models},
  author={Chu, Yunfei and Xu, Jin and Zhou, Xiaohuan and Yang, Qian and Zhang, Shiliang and Yan, Zhijie  and Zhou, Chang and Zhou, Jingren},
  journal={arXiv preprint arXiv:2311.07919},
  year={2023}
}

@article{Qwen2-Audio,
  title={Qwen2-Audio Technical Report},
  author={Chu, Yunfei and Xu, Jin and Yang, Qian and Wei, Haojie and Wei, Xipin and Guo,  Zhifang and Leng, Yichong and Lv, Yuanjun and He, Jinzheng and Lin, Junyang and Zhou, Chang and Zhou, Jingren},
  journal={arXiv preprint arXiv:2407.10759},
  year={2024}
}
}

\clearpage

\newpage
\section*{NeurIPS Paper Checklist}

\begin{enumerate}

\item {\bf Claims}
    \item[] Question: Do the main claims made in the abstract and introduction accurately reflect the paper's contributions and scope?
    \item[] Answer: \answerYes{} 
    \item[] Justification: The abstract explicitly states the core contributions and scope of this work.
    \item[] Guidelines:
    \begin{itemize}
        \item The answer NA means that the abstract and introduction do not include the claims made in the paper.
        \item The abstract and/or introduction should clearly state the claims made, including the contributions made in the paper and important assumptions and limitations. A No or NA answer to this question will not be perceived well by the reviewers. 
        \item The claims made should match theoretical and experimental results, and reflect how much the results can be expected to generalize to other settings. 
        \item It is fine to include aspirational goals as motivation as long as it is clear that these goals are not attained by the paper. 
    \end{itemize}

\item {\bf Limitations}
    \item[] Question: Does the paper discuss the limitations of the work performed by the authors?
    \item[] Answer: \answerYes{} 
    \item[] Justification: We include a "Limitation" section in the Appendix~\ref{appendix:limitations} to discuss the study's constraints.
    \item[] Guidelines:
    \begin{itemize}
        \item The answer NA means that the paper has no limitation while the answer No means that the paper has limitations, but those are not discussed in the paper. 
        \item The authors are encouraged to create a separate "Limitations" section in their paper.
        \item The paper should point out any strong assumptions and how robust the results are to violations of these assumptions (e.g., independence assumptions, noiseless settings, model well-specification, asymptotic approximations only holding locally). The authors should reflect on how these assumptions might be violated in practice and what the implications would be.
        \item The authors should reflect on the scope of the claims made, e.g., if the approach was only tested on a few datasets or with a few runs. In general, empirical results often depend on implicit assumptions, which should be articulated.
        \item The authors should reflect on the factors that influence the performance of the approach. For example, a facial recognition algorithm may perform poorly when image resolution is low or images are taken in low lighting. Or a speech-to-text system might not be used reliably to provide closed captions for online lectures because it fails to handle technical jargon.
        \item The authors should discuss the computational efficiency of the proposed algorithms and how they scale with dataset size.
        \item If applicable, the authors should discuss possible limitations of their approach to address problems of privacy and fairness.
        \item While the authors might fear that complete honesty about limitations might be used by reviewers as grounds for rejection, a worse outcome might be that reviewers discover limitations that aren't acknowledged in the paper. The authors should use their best judgment and recognize that individual actions in favor of transparency play an important role in developing norms that preserve the integrity of the community. Reviewers will be specifically instructed to not penalize honesty concerning limitations.
    \end{itemize}

\item {\bf Theory assumptions and proofs}
    \item[] Question: For each theoretical result, does the paper provide the full set of assumptions and a complete (and correct) proof?
    \item[] Answer: \answerNo{} 
    \item[] Justification: Our benchmark is primarily designed to evaluate model performance and does not involve theoretical assumptions or mathematical proofs.
    \item[] Guidelines:
    \begin{itemize}
        \item The answer NA means that the paper does not include theoretical results. 
        \item All the theorems, formulas, and proofs in the paper should be numbered and cross-referenced.
        \item All assumptions should be clearly stated or referenced in the statement of any theorems.
        \item The proofs can either appear in the main paper or the supplemental material, but if they appear in the supplemental material, the authors are encouraged to provide a short proof sketch to provide intuition. 
        \item Inversely, any informal proof provided in the core of the paper should be complemented by formal proofs provided in appendix or supplemental material.
        \item Theorems and Lemmas that the proof relies upon should be properly referenced. 
    \end{itemize}

    \item {\bf Experimental result reproducibility}
    \item[] Question: Does the paper fully disclose all the information needed to reproduce the main experimental results of the paper to the extent that it affects the main claims and/or conclusions of the paper (regardless of whether the code and data are provided or not)?
    \item[] Answer: \answerYes{} 
    \item[] Justification: We have shown the prompt templates as shown in Appendix~\ref{appendix:prompts}. We have also released prompt templates, evaluation code on \url{https://github.com/HychTus/BRACE_Evaluation} and benchmark data on \url{https://huggingface.co/datasets/gtysssp/audio_benchmarks} to ensure the reproducibility of our experiments.
    \item[] Guidelines:
    \begin{itemize}
        \item The answer NA means that the paper does not include experiments.
        \item If the paper includes experiments, a No answer to this question will not be perceived well by the reviewers: Making the paper reproducible is important, regardless of whether the code and data are provided or not.
        \item If the contribution is a dataset and/or model, the authors should describe the steps taken to make their results reproducible or verifiable. 
        \item Depending on the contribution, reproducibility can be accomplished in various ways. For example, if the contribution is a novel architecture, describing the architecture fully might suffice, or if the contribution is a specific model and empirical evaluation, it may be necessary to either make it possible for others to replicate the model with the same dataset, or provide access to the model. In general. releasing code and data is often one good way to accomplish this, but reproducibility can also be provided via detailed instructions for how to replicate the results, access to a hosted model (e.g., in the case of a large language model), releasing of a model checkpoint, or other means that are appropriate to the research performed.
        \item While NeurIPS does not require releasing code, the conference does require all submissions to provide some reasonable avenue for reproducibility, which may depend on the nature of the contribution. For example
        \begin{enumerate}
            \item If the contribution is primarily a new algorithm, the paper should make it clear how to reproduce that algorithm.
            \item If the contribution is primarily a new model architecture, the paper should describe the architecture clearly and fully.
            \item If the contribution is a new model (e.g., a large language model), then there should either be a way to access this model for reproducing the results or a way to reproduce the model (e.g., with an open-source dataset or instructions for how to construct the dataset).
            \item We recognize that reproducibility may be tricky in some cases, in which case authors are welcome to describe the particular way they provide for reproducibility. In the case of closed-source models, it may be that access to the model is limited in some way (e.g., to registered users), but it should be possible for other researchers to have some path to reproducing or verifying the results.
        \end{enumerate}
    \end{itemize}

\item {\bf Open access to data and code}
    \item[] Question: Does the paper provide open access to the data and code, with sufficient instructions to faithfully reproduce the main experimental results, as described in supplemental material?
    \item[] Answer: \answerYes{} 
    \item[] Justification: We have released our evaluation code and data in \url{https://github.com/HychTus/BRACE_Evaluation} and \url{https://huggingface.co/datasets/gtysssp/audio_benchmarks}.
    \item[] Guidelines:
    \begin{itemize}
        \item The answer NA means that paper does not include experiments requiring code.
        \item Please see the NeurIPS code and data submission guidelines (\url{https://nips.cc/public/guides/CodeSubmissionPolicy}) for more details.
        \item While we encourage the release of code and data, we understand that this might not be possible, so “No” is an acceptable answer. Papers cannot be rejected simply for not including code, unless this is central to the contribution (e.g., for a new open-source benchmark).
        \item The instructions should contain the exact command and environment needed to run to reproduce the results. See the NeurIPS code and data submission guidelines (\url{https://nips.cc/public/guides/CodeSubmissionPolicy}) for more details.
        \item The authors should provide instructions on data access and preparation, including how to access the raw data, preprocessed data, intermediate data, and generated data, etc.
        \item The authors should provide scripts to reproduce all experimental results for the new proposed method and baselines. If only a subset of experiments are reproducible, they should state which ones are omitted from the script and why.
        \item At submission time, to preserve anonymity, the authors should release anonymized versions (if applicable).
        \item Providing as much information as possible in supplemental material (appended to the paper) is recommended, but including URLs to data and code is permitted.
    \end{itemize}

\item {\bf Experimental setting/details}
    \item[] Question: Does the paper specify all the training and test details (e.g., data splits, hyperparameters, how they were chosen, type of optimizer, etc.) necessary to understand the results?
    \item[] Answer: \answerYes{} 
    \item[] Justification: All experimental configurations, including data construction and hyperparameters settings, are thoroughly documented in Section~\ref{sec: experiment_setting}.
    \item[] Guidelines:
    \begin{itemize}
        \item The answer NA means that the paper does not include experiments.
        \item The experimental setting should be presented in the core of the paper to a level of detail that is necessary to appreciate the results and make sense of them.
        \item The full details can be provided either with the code, in appendix, or as supplemental material.
    \end{itemize}

\item {\bf Experiment statistical significance}
    \item[] Question: Does the paper report error bars suitably and correctly defined or other appropriate information about the statistical significance of the experiments?
    \item[] Answer: \answerNo{} 
    \item[] Justification: Our benchmark focuses on evaluating LALMs and ACEMs for caption pair comparison tasks. We observe result variability stemming from two primary sources: (1)inherent stochasiticity in model behaviors; (2) variance in subjective rating protocols. We only calculate the mean and variance of CLAP models' results on BRACE, as shown in Table~\ref{tab:Performance-of-CLAPs-Updated}. For LALMs, computational constraints posed a practical limitation: each round of evaluation takes up to 14 hours, making it prohibitively expensive to conduct multiple runs for conventional error bar estimation.
    \item[] Guidelines:
    \begin{itemize}
        \item The answer NA means that the paper does not include experiments.
        \item The authors should answer "Yes" if the results are accompanied by error bars, confidence intervals, or statistical significance tests, at least for the experiments that support the main claims of the paper.
        \item The factors of variability that the error bars are capturing should be clearly stated (for example, train/test split, initialization, random drawing of some parameter, or overall run with given experimental conditions).
        \item The method for calculating the error bars should be explained (closed form formula, call to a library function, bootstrap, etc.)
        \item The assumptions made should be given (e.g., Normally distributed errors).
        \item It should be clear whether the error bar is the standard deviation or the standard error of the mean.
        \item It is OK to report 1-sigma error bars, but one should state it. The authors should preferably report a 2-sigma error bar than state that they have a 96\% CI, if the hypothesis of Normality of errors is not verified.
        \item For asymmetric distributions, the authors should be careful not to show in tables or figures symmetric error bars that would yield results that are out of range (e.g. negative error rates).
        \item If error bars are reported in tables or plots, The authors should explain in the text how they were calculated and reference the corresponding figures or tables in the text.
    \end{itemize}

\item {\bf Experiments compute resources}
    \item[] Question: For each experiment, does the paper provide sufficient information on the computer resources (type of compute workers, memory, time of execution) needed to reproduce the experiments?
    \item[] Answer: \answerYes{} 
    \item[] Justification: Seel Appendix~\ref{Appendix: brace evaluation strategies} for hardware details.
    \item[] Guidelines:
    \begin{itemize}
        \item The answer NA means that the paper does not include experiments.
        \item The paper should indicate the type of compute workers CPU or GPU, internal cluster, or cloud provider, including relevant memory and storage.
        \item The paper should provide the amount of compute required for each of the individual experimental runs as well as estimate the total compute. 
        \item The paper should disclose whether the full research project required more compute than the experiments reported in the paper (e.g., preliminary or failed experiments that didn't make it into the paper). 
    \end{itemize}
    
\item {\bf Code of ethics}
    \item[] Question: Does the research conducted in the paper conform, in every respect, with the NeurIPS Code of Ethics \url{https://neurips.cc/public/EthicsGuidelines}?
    \item[] Answer: \answerYes{} 
    \item[] Justification: Our benchmark data is derived from human-annotated datasets. While our benchmark data may inherit certain biases(see Appendix~\ref{appendix:ethics statement}), all code implementations adhere to NeurlIPS Ethics Guidelines. 
    \item[] Guidelines:
    \begin{itemize}
        \item The answer NA means that the authors have not reviewed the NeurIPS Code of Ethics.
        \item If the authors answer No, they should explain the special circumstances that require a deviation from the Code of Ethics.
        \item The authors should make sure to preserve anonymity (e.g., if there is a special consideration due to laws or regulations in their jurisdiction).
    \end{itemize}

\item {\bf Broader impacts}
    \item[] Question: Does the paper discuss both potential positive societal impacts and negative societal impacts of the work performed?
    \item[] Answer: \answerYes{} 
    \item[] Justification: See Appendix~\ref{appendix:broader impact} for detail.
    \item[] Guidelines:
    \begin{itemize}
        \item The answer NA means that there is no societal impact of the work performed.
        \item If the authors answer NA or No, they should explain why their work has no societal impact or why the paper does not address societal impact.
        \item Examples of negative societal impacts include potential malicious or unintended uses (e.g., disinformation, generating fake profiles, surveillance), fairness considerations (e.g., deployment of technologies that could make decisions that unfairly impact specific groups), privacy considerations, and security considerations.
        \item The conference expects that many papers will be foundational research and not tied to particular applications, let alone deployments. However, if there is a direct path to any negative applications, the authors should point it out. For example, it is legitimate to point out that an improvement in the quality of generative models could be used to generate deepfakes for disinformation. On the other hand, it is not needed to point out that a generic algorithm for optimizing neural networks could enable people to train models that generate Deepfakes faster.
        \item The authors should consider possible harms that could arise when the technology is being used as intended and functioning correctly, harms that could arise when the technology is being used as intended but gives incorrect results, and harms following from (intentional or unintentional) misuse of the technology.
        \item If there are negative societal impacts, the authors could also discuss possible mitigation strategies (e.g., gated release of models, providing defenses in addition to attacks, mechanisms for monitoring misuse, mechanisms to monitor how a system learns from feedback over time, improving the efficiency and accessibility of ML).
    \end{itemize}
    
\item {\bf Safeguards}
    \item[] Question: Does the paper describe safeguards that have been put in place for responsible release of data or models that have a high risk for misuse (e.g., pretrained language models, image generators, or scraped datasets)?
    \item[] Answer: \answerNo{} 
    \item[] Justification: The constructed dataset is specifically designed for evaluating performance of ACEMs and LALMs, with inherent safeguards against misuse: All content is derived from original open-source datasets and model-generated outputs, without involving any potential privacy concerns.
    \item[] Guidelines:
    \begin{itemize}
        \item The answer NA means that the paper poses no such risks.
        \item Released models that have a high risk for misuse or dual-use should be released with necessary safeguards to allow for controlled use of the model, for example by requiring that users adhere to usage guidelines or restrictions to access the model or implementing safety filters. 
        \item Datasets that have been scraped from the Internet could pose safety risks. The authors should describe how they avoided releasing unsafe images.
        \item We recognize that providing effective safeguards is challenging, and many papers do not require this, but we encourage authors to take this into account and make a best faith effort.
    \end{itemize}

\item {\bf Licenses for existing assets}
    \item[] Question: Are the creators or original owners of assets (e.g., code, data, models), used in the paper, properly credited and are the license and terms of use explicitly mentioned and properly respected?
    \item[] Answer: \answerYes{} 
    \item[] Justification: Our evaluation dataset is derived from AudioCaps (MIT License) and Clotho (Tampere University licence). All employed models strictly adhere to their respective usage agreements.
    \item[] Guidelines:
    \begin{itemize}
        \item The answer NA means that the paper does not use existing assets.
        \item The authors should cite the original paper that produced the code package or dataset.
        \item The authors should state which version of the asset is used and, if possible, include a URL.
        \item The name of the license (e.g., CC-BY 4.0) should be included for each asset.
        \item For scraped data from a particular source (e.g., website), the copyright and terms of service of that source should be provided.
        \item If assets are released, the license, copyright information, and terms of use in the package should be provided. For popular datasets, \url{paperswithcode.com/datasets} has curated licenses for some datasets. Their licensing guide can help determine the license of a dataset.
        \item For existing datasets that are re-packaged, both the original license and the license of the derived asset (if it has changed) should be provided.
        \item If this information is not available online, the authors are encouraged to reach out to the asset's creators.
    \end{itemize}

\item {\bf New assets}
    \item[] Question: Are new assets introduced in the paper well documented and is the documentation provided alongside the assets?
    \item[] Answer: \answerYes{} 
    \item[] Justification: We have provided the related materials following the submission guidelines.
    \item[] Guidelines:
    \begin{itemize}
        \item The answer NA means that the paper does not release new assets.
        \item Researchers should communicate the details of the dataset/code/model as part of their submissions via structured templates. This includes details about training, license, limitations, etc. 
        \item The paper should discuss whether and how consent was obtained from people whose asset is used.
        \item At submission time, remember to anonymize your assets (if applicable). You can either create an anonymized URL or include an anonymized zip file.
    \end{itemize}

\item {\bf Crowdsourcing and research with human subjects}
    \item[] Question: For crowdsourcing experiments and research with human subjects, does the paper include the full text of instructions given to participants and screenshots, if applicable, as well as details about compensation (if any)? 
    \item[] Answer: \answerYes{} 
    \item[] Justification: Limited human evaluation was conducted soely for benchmark annotation.
    \item[] Guidelines:
    \begin{itemize}
        \item The answer NA means that the paper does not involve crowdsourcing nor research with human subjects.
        \item Including this information in the supplemental material is fine, but if the main contribution of the paper involves human subjects, then as much detail as possible should be included in the main paper. 
        \item According to the NeurIPS Code of Ethics, workers involved in data collection, curation, or other labor should be paid at least the minimum wage in the country of the data collector. 
    \end{itemize}

\item {\bf Institutional review board (IRB) approvals or equivalent for research with human subjects}
    \item[] Question: Does the paper describe potential risks incurred by study participants, whether such risks were disclosed to the subjects, and whether Institutional Review Board (IRB) approvals (or an equivalent approval/review based on the requirements of your country or institution) were obtained?
    \item[] Answer: \answerNo{} 
    \item[] Justification: Human annotators were limited to non-interventional annotation tasks conducted internallly (institutionally exempt, no external participants or personal data collected).
    \item[] Guidelines:
    \begin{itemize}
        \item The answer NA means that the paper does not involve crowdsourcing nor research with human subjects.
        \item Depending on the country in which research is conducted, IRB approval (or equivalent) may be required for any human subjects research. If you obtained IRB approval, you should clearly state this in the paper. 
        \item We recognize that the procedures for this may vary significantly between institutions and locations, and we expect authors to adhere to the NeurIPS Code of Ethics and the guidelines for their institution. 
        \item For initial submissions, do not include any information that would break anonymity (if applicable), such as the institution conducting the review.
    \end{itemize}

\item {\bf Declaration of LLM usage}
    \item[] Question: Does the paper describe the usage of LLMs if it is an important, original, or non-standard component of the core methods in this research? Note that if the LLM is used only for writing, editing, or formatting purposes and does not impact the core methodology, scientific rigorousness, or originality of the research, declaration is not required.
    \item[] Answer: \answerYes{} 
    \item[] Justification: As show in Section~\ref{sec:brace-hallucination benchmark} and Section~\ref{sec:brace-main benchmark}, we leverage LLMs for filtering raw data and generating hallucinated synthetic samples. Furthermore, LALMs are applied to produce audio captions. 
    \item[] Guidelines:
    \begin{itemize}
        \item The answer NA means that the core method development in this research does not involve LLMs as any important, original, or non-standard components.
        \item Please refer to our LLM policy (\url{https://neurips.cc/Conferences/2025/LLM}) for what should or should not be described.
    \end{itemize}

\end{enumerate}

\appendix



\newpage
\section{BRACE evaluation strategies} \label{Appendix: brace evaluation strategies}
\begin{figure}[h]
  \centering
  \includegraphics[width=\textwidth]{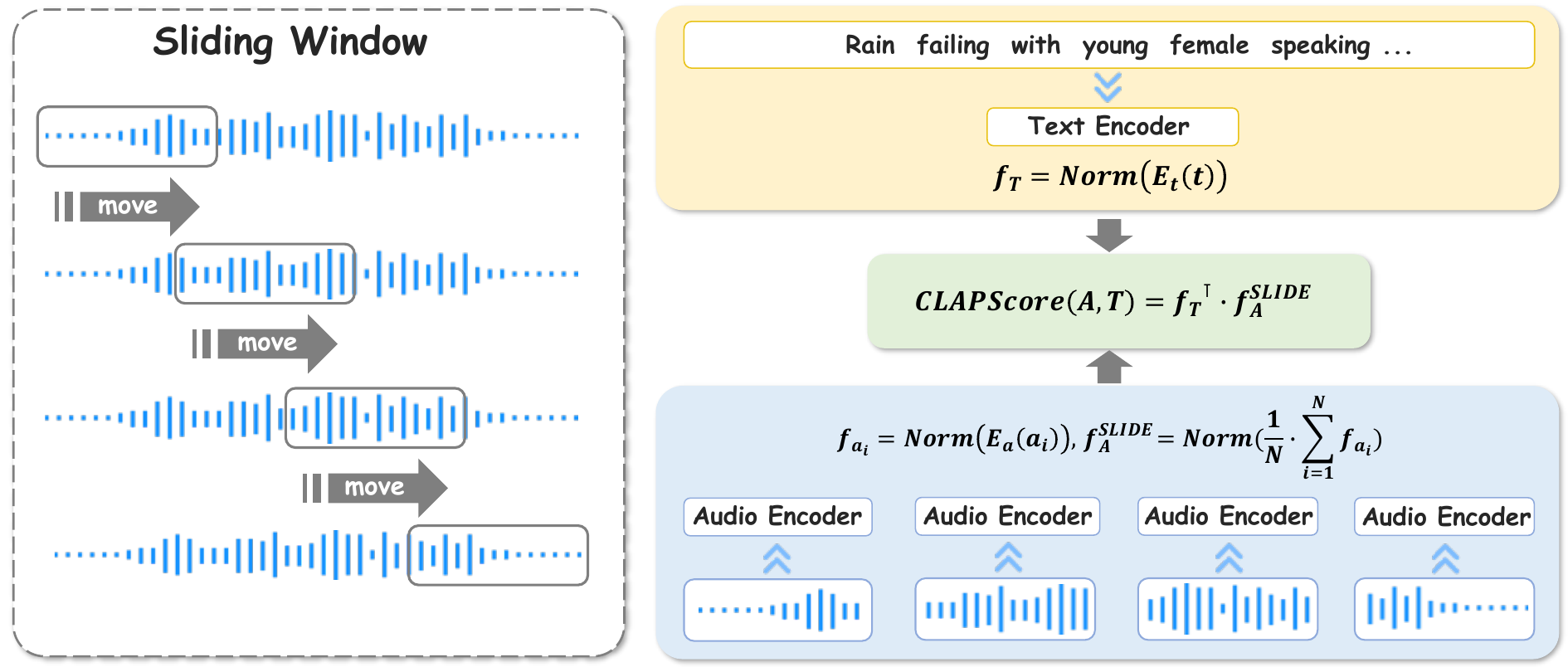} 
  \caption{We introduce a sliding window strategy, where the CLAP embeddings from each audio segment are averaged across overlapping windows. This approach allows the model to effectively capture representations from the entire audio clip, even when its duration exceeds the fixed input window length of the CLAP encoder. By aggregating embeddings across all segments, the final audio representation retains global contextual information and can be reliably used for similarity computation with the corresponding caption. Moreover, this method also addresses the reproducibility issue caused by the CLAP encoder’s random window truncation when the input audio length exceeds the fixed window size.}
  \label{fig: slide}
\end{figure}


As BRACE serves different purposes for CLAPScore and LALM evaluation, we adopt distinct evaluation strategies tailored to each. In the following sections, we detail its configuration and evaluation procedure. To ensure that experimental results are both reproducible and comparable, we standardize the evaluation strategies for each model type accordingly. For all the experiments, we evaluate models on 8 × NVIDIA H100 64G.

\subsection{CLAP evaluation strategy}

Given an audio sample $A$ and a caption $T$, the CLAP model encodes the audio and text independently using its audio encoder $E_a$ and text encoder $E_t$. The resulting embeddings, $\mathbf{f}_A$ and $\mathbf{f}_T$, are then normalized. The similarity between the audio and the caption is computed using the dot product: $\mathbf{f}_T^{\top} \cdot \mathbf{f}_A$.

To evaluate CLAP on the BRACE benchmark, we compute similarity scores between the audio $A$ and two captions, $T_0$ and $T_1$, resulting in CLAPScores $score_0$ and $score_1$, respectively. These scores are interpreted as preference indicators. 

The model’s preference is determined using the following decision rule:


\begin{equation}
    \text{CLAP-Preference} = 
    \begin{cases}
        \text{caption}_0, & \text{if } score_0 \ge score_1 \\
        \text{caption}_1, & \text{if } score_1 \ge score_0 \\
    \end{cases}
\end{equation}

This computed preference, referred to as CLAP-Preference, is then compared against human annotations to assess how well the model's judgments align with human perception.  
Since CLAP encodes fixed-length audio segments, shorter inputs are padded and repeated while longer ones are truncated, which introduces randomness into the evaluation. As a result, the outputs may vary across runs, affecting consistency. To address this limitation, we propose the SLIDE-CLAP strategy.

\subsection{SLIDE-CLAP evaluation strategy}

To reduce randomness and improve reproducibility, we adopt a sliding window strategy for computing audio embeddings. As shown in Figure~\ref{fig: slide}, we use the model’s inherent fixed-length audio encoding size as the window size, and all models share the same hop size. The raw audio $A$ is segmented using a sliding window into $N$ audio clips $\{a_i\}_{i=1}^{N}$. Each clip is encoded using the audio encoder to obtain embeddings $f_{a_i}$, and the final audio embedding is computed as the average of these embeddings, as shown in the following equation:
\begin{equation}
    \begin{aligned}
        &\mathbf{f}_{a_i} = \text{Norm} (E_a(a_i)), 
        \quad \mathbf{f}_{A}^{\text{SLIDE}} = \text{Norm} \left( \frac{1}{N} \cdot \sum_{i=1}^{N}  \mathbf{f}_{a_i} \right) \\
        &\text{SLIDE-CLAP-Score}(A, T) = \mathbf{f}_{T}^{\top}  \cdot \mathbf{f}_{A}^{\text{SLIDE}}
    \end{aligned}
\end{equation}

Apart from the way the audio embedding is computed, all other evaluation procedures remain the same as in the original CLAP strategy.  

\subsection{LALM evaluation strategy} \label{appendix:lalm evaluation stragtegy}

\paragraph{LALM prompt setting} We embed both \texttt{caption\_0} and \texttt{caption\_1} into a standardized prompt template, instructing the model to identify the superior caption based on the given audio input. Due to the inherent instability of LALMs output for different prompts, we utilize a diverse set of prompt templates across all LALMs and report the best results. These templates range from simple to more sophisticated designs. 

We define explicit \textbf{evaluation criteria}, guiding the model in evaluating aspects such as alignment of entities, consistency of events, avoidance of hallucinations, and linguistic quality. For all non-naive prompts, the task is presented in a \textbf{multiple-choice format}, with clear instructions for the model to select a single preferred option. Additionally, we design prompts both with and without a "tie" option, and with or without access to reference captions, to facilitate a more comprehensive evaluation of model behavior. All prompt variants are detailed in Appendix~\ref{appendix:prompts}.

\paragraph{LALM output processing} Since most existing LALMs are instruction-tuned for open-ended generation and exhibit limited instruction-following capability, we do not impose a rigid output format. Instead, we employ a text-based language model with stronger instruction adherence to distill the LALM’s output into a final summarized preference. The prompt used for this secondary model is also provided in Figure~\ref{fig: summary_prompt}. The distilled preference is categorized as one of the following: \texttt{caption\_0}, \texttt{caption\_1}, \texttt{tie}, or \texttt{unknown}, with unknown indicating that the model could not infer a definitive preference from the LALM’s response.

\paragraph{LALM result calculation} Since the benchmark data has been curated to ensure clearly distinguishable preferences, any ambiguous outputs are treated as incorrect. Specifically, predictions are marked incorrect if LALM outputs a tie or unknown.

\section{Addtional results}

\subsection{LALM results analysis}

Table~\ref{tab:lalm-hallu-distribution} reveals that prompts have a noticeable yet seemingly random effect on the output distribution of LALMs. There is no clear or consistent trend in model behavior across prompt types—from naive to simple to complex—suggesting a high degree of instability in how LALMs respond to varying prompt structures.

Moreover, the models demonstrate limited comprehension abilities. In many cases, they fail to correctly interpret the instructions provided in the prompt, leading to outputs that do not reflect any meaningful preference. This is particularly evident in the frequent selection of the "Unknown" option, which indicates the model’s inability to engage with the task effectively. Notably, the LTU model selected "Unknown" in 47.84\% of cases under the naive + non-tie prompt setting, underscoring this issue.

In addition, a strong position bias persists across several models—most prominently in GAMA—which tend to favor the "Zero" or "One" option disproportionately, regardless of content relevance. This suggests that these models often rely on positional heuristics learned during training rather than genuine understanding of the input, further limiting their reasoning capability.

Overall, these findings point to significant challenges in the robustness, interpretability, and reasoning consistency of current LALMs.

\begin{table}[ht]
\caption{Output distribution of LALMs on BRACE-Hallucination}
\label{tab:lalm-hallu-distribution}
\centering
\setlength{\tabcolsep}{3.5pt}

\begin{tabular}{l*{7}{c}}  
\toprule
& \multicolumn{3}{c}{\textbf{Non-Tie}} & \multicolumn{4}{c}{\textbf{Tie}} \\
\cmidrule(lr){2-4} \cmidrule(lr){5-8}
\textbf{Model} & \textbf{Zero} & \textbf{One} & \textbf{Unknown} & \textbf{Zero} & \textbf{One} & \textbf{Tie} & \textbf{Unknown} \\
\midrule
\multicolumn{8}{c}{\textbf{Naive}} \\
\midrule
AF2 & 31.15 & 61.54 & 7.31 & 23.39 & 68.95 & 0.14 & 7.52 \\
LTU & 45.41 & 6.74 & 47.84 & 25.49 & 5.49 & 24.82 & 44.21 \\
GAMA & 86.56 & 6.61 & 6.84 & 90.7 & 3.11 & 0.67 & 5.52 \\
Qwen-Audio-Chat & 77.82 & 13.84 & 8.34 & 84.18 & 8.22 & 0.08 & 7.52 \\
Qwen2-Audio-Instruct & 36.62 & 23.14 & 40.24 & 27.35 & 14.88 & 12.8 & 44.96 \\
\midrule
\multicolumn{8}{c}{\textbf{Simple}} \\
\midrule
AF2 & 90.07 & 9.84 & 0.1 & 91.66 & 8.31 & 0.0 & 0.03 \\
LTU & 24.23 & 72.09 & 3.68 & 28.2 & 41.12 & 28.0 & 2.68 \\
GAMA & 96.52 & 0.5 & 2.98 & 97.0 & 0.13 & 0.15 & 2.71 \\
Qwen-Audio-Chat & 42.91 & 54.15 & 2.94 & 44.99 & 52.94 & 0.07 & 2.0 \\
Qwen2-Audio-Instruct & 61.76 & 28.22 & 10.02 & 49.1 & 27.21 & 21.95 & 1.74 \\
\midrule
\multicolumn{8}{c}{\textbf{Complex}} \\
\midrule
AF2 & 99.7 & 0.18 & 0.12 & 62.54 & 0.37 & 0.02 & 37.07 \\
LTU & 95.26 & 3.97 & 0.77 & 56.69 & 42.07 & 0.03 & 1.21 \\
GAMA & 89.62 & 0.22 & 10.16 & 96.31 & 0.01 & 0.0 & 3.68 \\
Qwen-Audio-Chat & 64.85 & 32.9 & 2.25 & 65.99 & 32.18 & 0.02 & 1.81 \\
Qwen2-Audio-Instruct & 68.18 & 30.13 & 1.68 & 37.49 & 2.77 & 58.79 & 0.94 \\
\bottomrule
\end{tabular}
\vspace{-4mm}
\end{table}

\subsection{CLAP results analysis}

As shown in Table~\ref{tab:Performance-of-CLAPs-Updated}, CLAP models exhibit significant variability across runs. For instance, MS-CLAP-2023 on AudioCaps shows a standard deviation of 1.06 and a range of F1-score from 51.02 to 56.32. This instability is caused by the fixed input window, which requires random cropping of long audio clips. Such randomness introduces inconsistency in the results and affects reproducibility.

Table~\ref{tab:Detail-Main-Results-on-BRACE-Main} shows that SLIDE-CLAP models perform best on Human-Machine (HM) pairs—for example, LAION-CLAP achieves 81.09 on HM2—while struggling with Human-Human (HH) and Machine-Machine (MM) comparisons, where scores generally fall below 70. This suggests that current models are better at detecting large stylistic gaps than subtle quality differences, highlighting limitations in fine-grained caption evaluation.

Due to the randomness in the audio window truncation by CLAP models, we use sliding window mechanism as our default setting and refer to SLIDE-CLAP as CLAP in the subsequent chapters of the appendix for brevity.

\begin{table}[ht]
  \vspace{-4mm}
  \caption{Statistical analysis of CLAP's results on BRACE-Main and BRACE-Hallucination. The table presents the mean, standard deviation, minimum, and maximum values for each model across AudioCaps and Clotho. The data presented in the table represents the results produced by the CLAP models over twenty independent experimental runs.}
  \label{tab:Performance-of-CLAPs-Updated}
  \centering
  \begin{tabular}{lcccccccc}
    \toprule
    & \multicolumn{4}{c}{\textbf{AudioCaps}} & \multicolumn{4}{c}{\textbf{Clotho}} \\
    \cmidrule(lr){2-5} \cmidrule(lr){6-9}
    \textbf{Model} & mean & std & min & max & mean & std & min & max \\
    \midrule
    \multicolumn{9}{c}{\textbf{BRACE-Main}} \\
    \midrule
    M2D-CLAP  & 62.96 & 0.60 & 61.69 & 64.31 & 56.61 & 0.78 & 54.68 & 58.74 \\
    MS-CLAP-2022 & 54.93 & 0.95 & 51.61 & 57.09 & 69.13 & 0.79 & 66.82 & 71.06 \\
    MS-CLAP-2023 & 53.56 & 1.06 & 51.02 & 56.32 & 67.58 & 0.93 & 65.08 & 70.25 \\
    LAION-CLAP & 73.33 & 0.62 & 71.86 & 74.53 & 64.54 & 0.80 & 62.32 & 66.97 \\
    \midrule
    \multicolumn{9}{c}{\textbf{BRACE-Hallucination}} \\
    \midrule
    M2D-CLAP  & 90.47 & 0.28 & 89.91 & 91.33 & 81.91 & 0.38 & 81.18 & 83.02 \\
    MS-CLAP-2022 & 74.43 & 0.57 & 73.08 & 75.90 & 88.66 & 0.34 & 87.72 & 89.39 \\
    MS-CLAP-2023 & 79.15 & 0.56 & 77.71 & 80.54 & 83.45 & 0.41 & 82.42 & 84.41 \\
    LAION-CLAP & 86.99 & 0.36 & 86.25 & 88.01 & 78.88 & 0.35 & 78.09 & 79.91 \\
    \bottomrule
  \end{tabular}
\end{table}


\begin{table}[ht]
  \caption{Detailed performance of SLIDE-CLAPs on BRACE-Main. Results are shown across different caption pair types. Detailed information about different pair types is
  shown in Table~\ref{table: six pairs and explanation}.}
  \label{tab:Detail-Main-Results-on-BRACE-Main}
  \centering
  \begin{tabular}{lcccccc}
    \toprule
    \textbf{Model} & HH & HM1 & HM2 & MM1 & MM2 & MM3 \\
    \midrule
    M2D-CLAP          & 49.23           & 61.83          & 66.40            & 64.34          & 53.96          & \textbf{69.16} \\
    MS-CLAP-2022      & 54.42           & 55.58          & 74.76            & 58.27          & \textbf{65.71} & 57.28 \\
    MS-CLAP-2023      & \textbf{63.21}  & 60.00          & 67.73            & 52.28          & 60.74          & 60.20 \\
    LAION-CLAP        & 57.58           & \textbf{75.22} & \textbf{81.09}   & \textbf{66.63} & 63.24          & 67.40 \\
    \bottomrule
  \end{tabular}
\end{table}

\subsection{Evaluation LALM and CLAP with references}
When incorporating reference captions, we observe different behaviors between CLAP models and LALMs. 
\paragraph{CLAP with references} Table~\ref{tab:clap-main-and-hallu-ref} shows the performance of CLAP models using different numbers of reference captions. For BRACE-Main, adding references leads to a noticeable improvement, where models that originally performed poorly see significant gains. This suggests that most CLAP models struggle with the alignment between text and audio modalities, and they benefit from reference captions to better distinguish between two captions. For BRACE-Hallucination, the inclusion of references further enhances the ability to detect hallucinations, indicating that reference captions provide clearer signals for identifying hallucinated content in captions.

\begin{table}[ht]
\vspace{-2mm}
\caption{Performance of CLAP models on BRACE-Main and BRACE-Hallucination using different numbers of references.}
\label{tab:clap-main-and-hallu-ref}
\centering
\setlength{\tabcolsep}{5pt}
\begin{tabular}{l*{9}{c}}
\toprule
& \multicolumn{3}{c}{\textbf{AudioCaps}} & \multicolumn{3}{c}{\textbf{Clotho}} & \multicolumn{3}{c}{\textbf{Avg}} \\
\cmidrule(lr){2-4} \cmidrule(lr){5-7} \cmidrule(lr){8-10}
\textbf{Model} & \textbf{1 Ref} & \textbf{3 Refs} & \textbf{5 Refs} & \textbf{1 Ref} & \textbf{3 Refs} & \textbf{5 Refs} & \textbf{1 Ref} & \textbf{3 Refs} & \textbf{5 Refs} \\
\midrule
\multicolumn{10}{c}{\textbf{BRACE-Main}} \\
\midrule
M2D-CLAP & 68.29 & 71.58 & 71.39 & 65.38 & 69.36 & 70.27 & 66.84 & 70.47 & 70.83 \\
MS-CLAP-2022 & 68.57 & 69.14 & 69.78 & \textbf{71.26} & \textbf{72.45} & \textbf{72.96} & 69.92 & 70.80 & 71.37 \\
MS-CLAP-2023 & 70.26 & 70.90 & 71.31 & 68.22 & 72.31 & 72.71 & 69.24 & 71.61 & 72.01 \\
LAION-CLAP & \textbf{75.02} & \textbf{76.10} & \textbf{76.14} & 68.05 & 68.93 & 69.35 & \textbf{71.54} & \textbf{72.52} & \textbf{72.74} \\
\midrule
\multicolumn{10}{c}{\textbf{BRACE-Hallucination}} \\
\midrule
M2D-CLAP & \textbf{97.13} & \textbf{97.92} & \textbf{97.84} & \textbf{94.58} & \textbf{96.91} & \textbf{98.00} & \textbf{95.85} & \textbf{97.41} & \textbf{97.92} \\
MS-CLAP-2022 & 91.94 & 93.72 & 93.79 & 94.44 & 95.76 & 96.39 & 93.19 & 94.74 & 95.09 \\
MS-CLAP-2023 & 94.18 & 96.01 & 96.37 & 92.58 & 94.55 & 95.24 & 93.38 & 95.28 & 95.80 \\
LAION-CLAP & 93.84 & 95.67 & 96.00 & 87.79 & 90.81 & 92.10 & 90.81 & 93.24 & 94.05 \\
\bottomrule
\end{tabular}
\vspace{-3mm}
\end{table}

\paragraph{LALM with references} On the other hand, most LALMs perform worse with references from Table \ref{tab:lalm-main-and-hallu-ref} due to poor instruction-following abilities. They struggle with the additional information and sometimes produce incorrect answers, such as selecting the reference caption as the better one. However, a few models like Qwen2-Audio-Instruct show significant improvement with references, achieving 64.24 on BRACE-Main and 79.00 on BRACE-Hallucination. Overall, while some models benefit, most LALMs are hindered by references rather than helped, underlining their general weakness in following instructions and understanding context.

\begin{table}[ht]
\centering
\vspace{-2mm}
\caption{Performance of LALMs on BRACE-Main and BRACE-Hallucination using different prompt templates with single reference.}
\label{tab:lalm-main-and-hallu-ref}
\begin{tabular}{l*{6}{c}}
\toprule
& \multicolumn{3}{c}{\textbf{Non-Tie}} & \multicolumn{3}{c}{\textbf{Tie}} \\
\cmidrule(lr){2-4} \cmidrule(lr){5-7}
\textbf{Model} & \textbf{Naive} & \textbf{Simple} & \textbf{Complex} & \textbf{Naive} & \textbf{Simple} & \textbf{Complex} \\
\midrule
\multicolumn{7}{c}{\textbf{BRACE-Main}} \\
\midrule
AF2 & 32.10 & 23.63 & 1.30 & 37.90 & \textbf{37.29} & 3.78 \\
LTU & \textbf{20.62} & 0.72 & 0.00 & 17.27 & 0.18 & 0.00 \\
GAMA & \textbf{25.84} & 23.68 & 25.11 & 9.48 & 5.25 & 0.36 \\
Qwen-Audio-Chat & 38.29 & \textbf{55.33} & 50.99 & 39.92 & 54.21 & 44.72 \\
Qwen2-Audio-Instruct & 49.52 & 63.74 & 61.35 & 41.46 & \textbf{64.24} & 29.93 \\
\midrule
\multicolumn{7}{c}{\textbf{BRACE-Hallucination}} \\
\midrule
AF2 & 51.45 & 23.55 & 0.86 & \textbf{56.32} & 40.11 & 1.25 \\
LTU & 13.34 & 1.97 & 0.00 & 10.39 & 0.24 & 0.00 \\
GAMA & \textbf{29.92} & 22.69 & 26.99 & 11.33 & 4.27 & 0.95 \\
Qwen-Audio-Chat & 51.45 & \textbf{60.20} & 55.29 & 55.68 & 58.60 & 41.27 \\
Qwen2-Audio-Instruct & 62.23 & \textbf{79.00} & 64.66 & 48.42 & 78.30 & 17.64 \\
\bottomrule
\end{tabular}
\end{table}

\clearpage
\section{Comprehensive data analysis}

\begin{table}[h]
    \caption{As shown in this table, we present three types of audio-caption pairs for each audio clip in our BRACE-Main benchmark.  
    \textit{human} stands for the human-annotated captions of a audio,  
    \textit{generation} stands for captions generated by LTU or GAMA for a audio clip, whereas \textit{corruption} stands for captions generated by models after corruption. \textit{H} stands for human-annotated captions, while \textit{M} represents machine-generated captions. The HM and MM categories are further subdivided into additional subcategories for more granular comparison. HM1 refers to human-annotated captions paired with captions generated by LTU or GAMA. HM2 represents human-annotated captions paired with captions corrupted by large language models. MM1 denotes captions generated by models paired with captions from different models. MM2 represents machine-generated captions paired with corrupted captions. MM3 involves corrupted captions paired with other corrupted captions.}
    \begin{center}
    \setlength{\tabcolsep}{28pt}
    \begin{tabular}{ccc}
    \toprule
    Pair Groups                     & Caption 1 & Caption 2  \\
    \midrule
    \textit{HH} & human & human\\
    \textit{HM1} & human & generated \\
    \textit{HM2} & human & corrupted \\ 
    \textit{MM1}         & generated & generated \\
    \textit{MM2}         & generated & corrupted \\
    \textit{MM3}         & corrupted & corrupted \\
    \bottomrule
    \end{tabular}
    \end{center}
    \label{table: six pairs and explanation}
    \vspace{4mm}
\end{table}

\begin{table}[h]
\vspace{-4mm}
\caption{Comparison of Fleiss' Kappa Scores Before and After Data Filtering. Fleiss’ Kappa is a statistical measure used to evaluate inter-annotator agreement. A significant improvement in the score after data filtering indicates increased annotation consistency, thereby reflecting the enhanced quality and reliability of our benchmark.}
\begin{center}
\setlength{\tabcolsep}{28pt}
\begin{tabular}{cccc}
\toprule
Dataset                 & AudioCaps & Clotho  \\
\midrule
Before Filtering         & 0.3806 & 0.4380            \\
After Filtering          & 0.9822 & 0.8422            \\
\bottomrule
\end{tabular}
\label{Fleiss-Kappa-table}
\end{center}
\vspace{-3.5mm}
\end{table}

\begin{figure}[h]
    \centering
    \includegraphics[width=\textwidth]{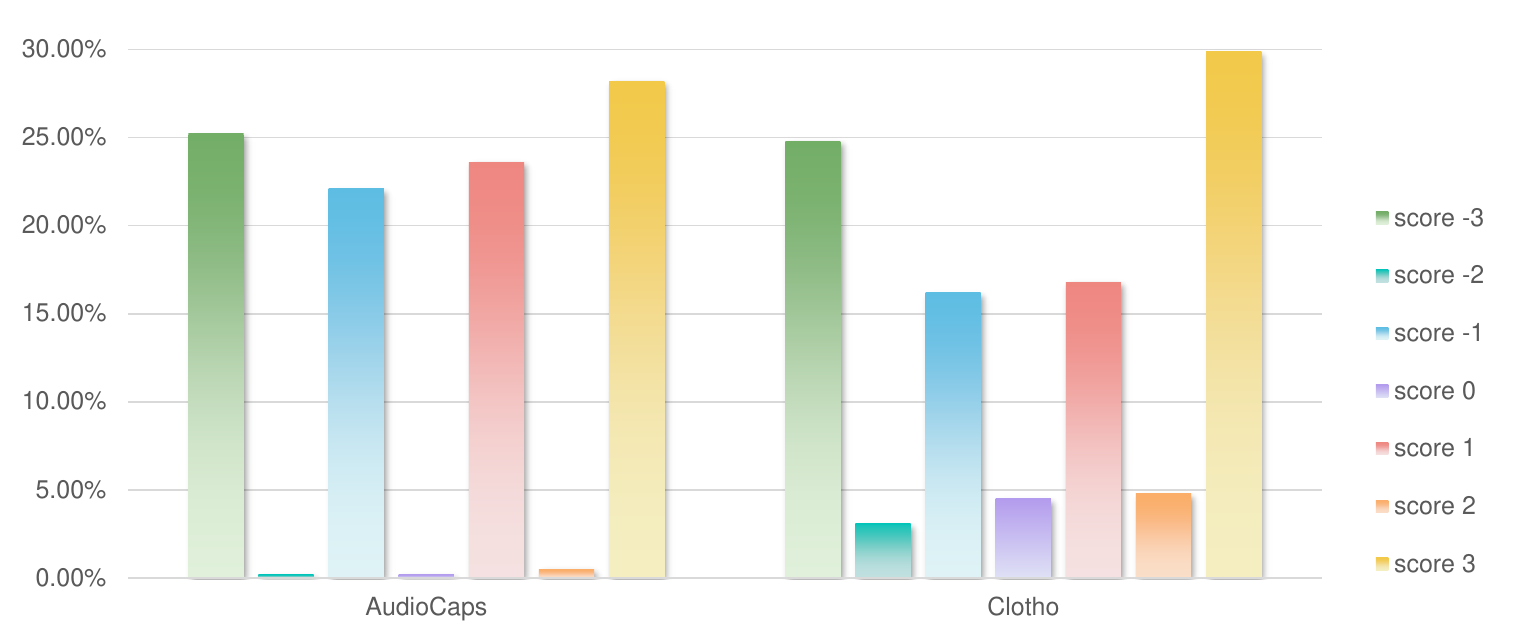}
    \caption{Distribution of Scores by Dataset. A clear variance is observed across different datasets. Notably, lower absolute scores indicate greater disagreement among human annotators, reflecting lower inter-annotator agreement. Such annotations are less reliable and therefore may not be suitable for constructing a high-quality benchmark.}
    \label{fig: Distribution of Scores by Dataset}
\end{figure}






\clearpage
\section{Case study}
\label{case-study}

\subsection{CLAP cases}
 MS-CLAP-2022 and MS-CLAP-2023 tend to overlook grammatical issues in the caption and did not recognize that caption\_0 of \texttt{106126.wav} only provided a partial description. Examples \texttt{Storm coming.wav} and \texttt{International Harvester Scout II.wav} demonstrate that the models overlook background sounds like traffic, leading to a loss of acoustic information. Example \texttt{CNC Machine 02.wav} shows that MS-CLAP-2022 and M2D-CLAP mistook a "whirring" sound for a "siren". \texttt{Hang Man\&39s Rope.wav} reveals models like MS-CLAP-2023 associating the sound with a "squirrel" instead of a "person". "Chopping Celery.wav" highlights errors in material identification, mistaking "metal" for "plastic". These cases show CLAP models' syntactic oversight and fine-grained acoustic perception issues, indicating areas for improvement.

\begin{figure}[htbp]
    \centering
    \begin{minipage}[t]{0.49\textwidth}
        \centering
        \includegraphics[width=\textwidth]{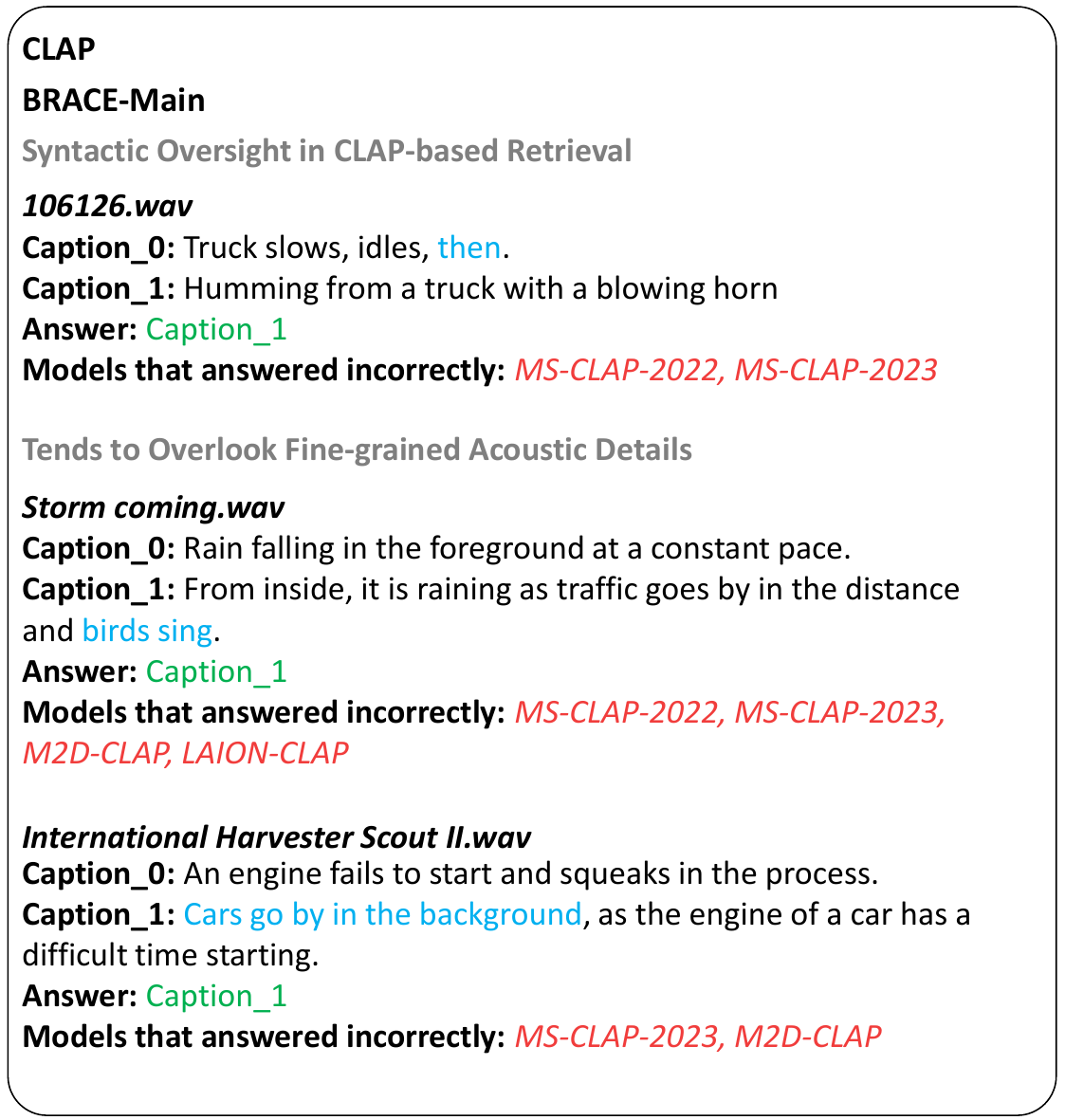}
        \label{fig:examples_clap_appendix_1}
    \end{minipage}
    \hfill
    \begin{minipage}[t]{0.49\textwidth}
        \centering
        \includegraphics[width=\textwidth]{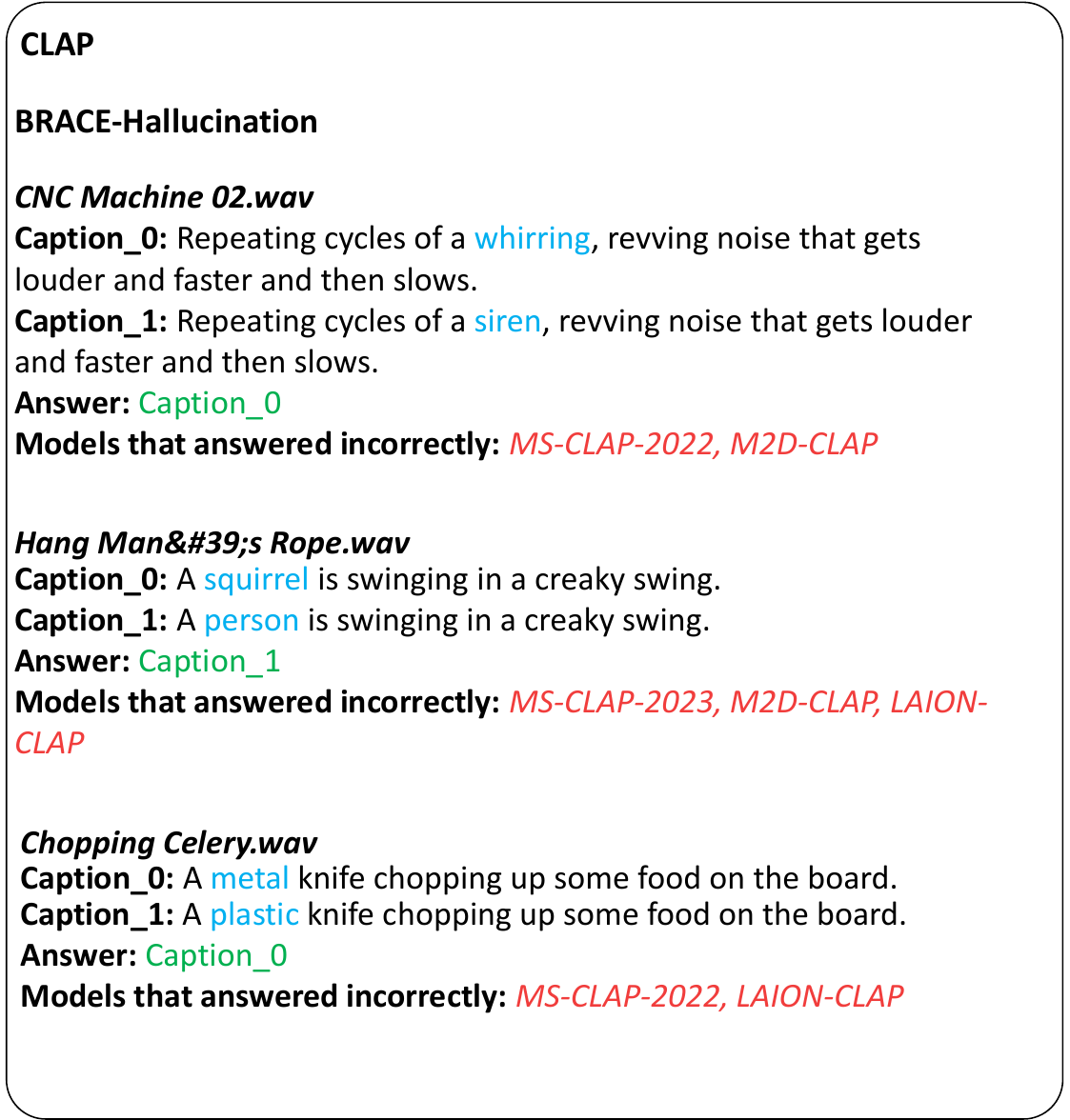}
        \label{fig:examples_clap_appendix_2}
    \end{minipage}
    \caption{Failure Cases of the CLAP Models on the BRACE Benchmark}
    \label{fig:examples_clap_appendix}
\end{figure}

\subsection{LALM cases}

\begin{figure}[h]
    \centering
    \includegraphics[width=\textwidth]{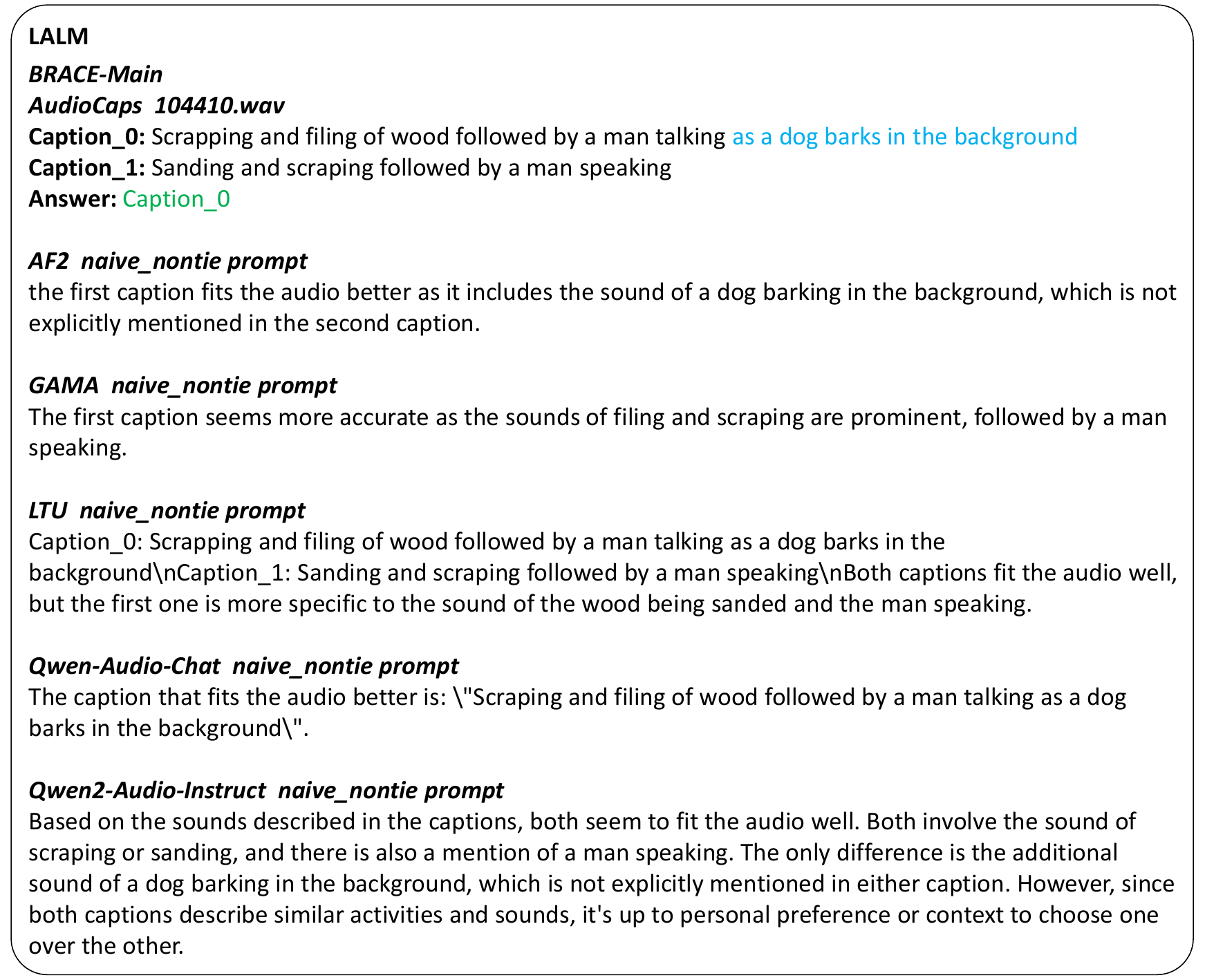}
    \caption{Cases of the LALMs on the BRACE-Main Benchmark}
    \label{fig:examples_lalms_appendix_3}
\end{figure}

\begin{figure}[h]
    \centering
    \includegraphics[width=\textwidth]{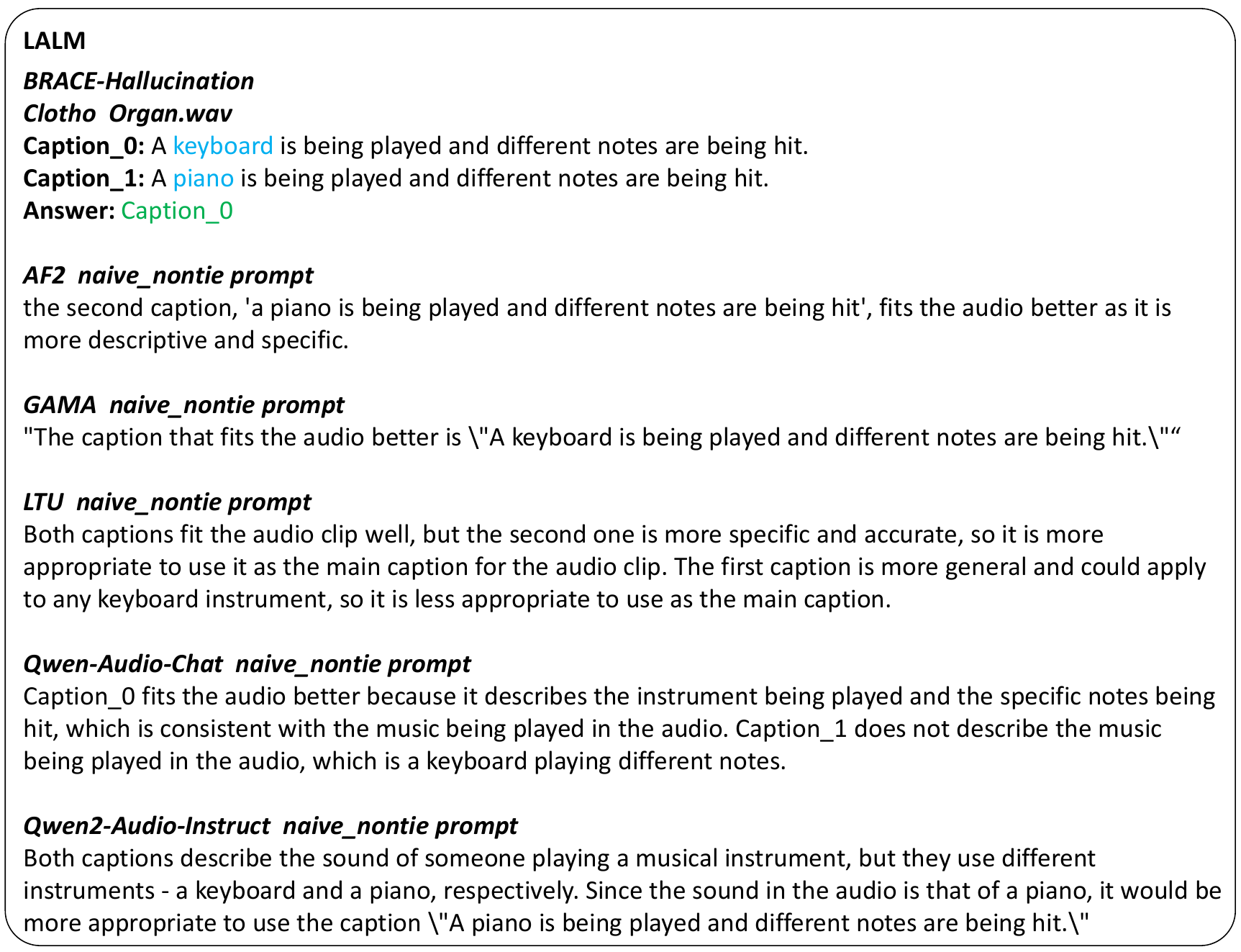}
    \caption{Cases of the LALMs on the BRACE-Hallucination Benchmark}
    \label{fig:examples_lalms_appendix_1}
\end{figure}

Figures~\ref{fig:examples_lalms_appendix_3} and~\ref{fig:examples_lalms_appendix_1} showcase the outputs of different LALMs given the same question and the same prompt template (\texttt{naive\_nontie}). The observed cases highlight several limitations in current models regarding instruction following, audio perception, and audio-text alignment. A detailed analysis of the responses from each model in Figure~\ref{fig:examples_lalms_appendix_1} is presented as follows:

\begin{itemize}
\item \textbf{AF2}: While this model attempts to provide reasoning, its justifications are vague and largely based on the caption text alone, lacking concrete connections to the audio modality.

\item \textbf{GAMA}: Exhibits a strong position bias, often defaulting to the first caption without substantive reasoning. This behavior undermines its ability to differentiate based on semantic content or audio cues.

\item \textbf{LTU}: Demonstrates poor language understanding and fails to capture the key semantic distinctions between the captions. The model is unable to detect fine-grained hallucinations, highlighting limitations in both text processing and multi-modal reasoning.

\item \textbf{Qwen-Audio-Chat}: This model shows comparatively strong multimodal capabilities. It successfully identifies the hallucination in the BRACE-caption—specifically, the substitution of “keyboard” with “piano”—and correctly cross-references with the audio to justify its choice.

\item \textbf{Qwen2-Audio-Instruct}: Although this model detects the textual differences between captions, it misjudges the correct alignment with the audio, indicating weaknesses in audio perception or integration.
\end{itemize}

\begin{figure}[h]
    \centering
    \includegraphics[width=\textwidth]{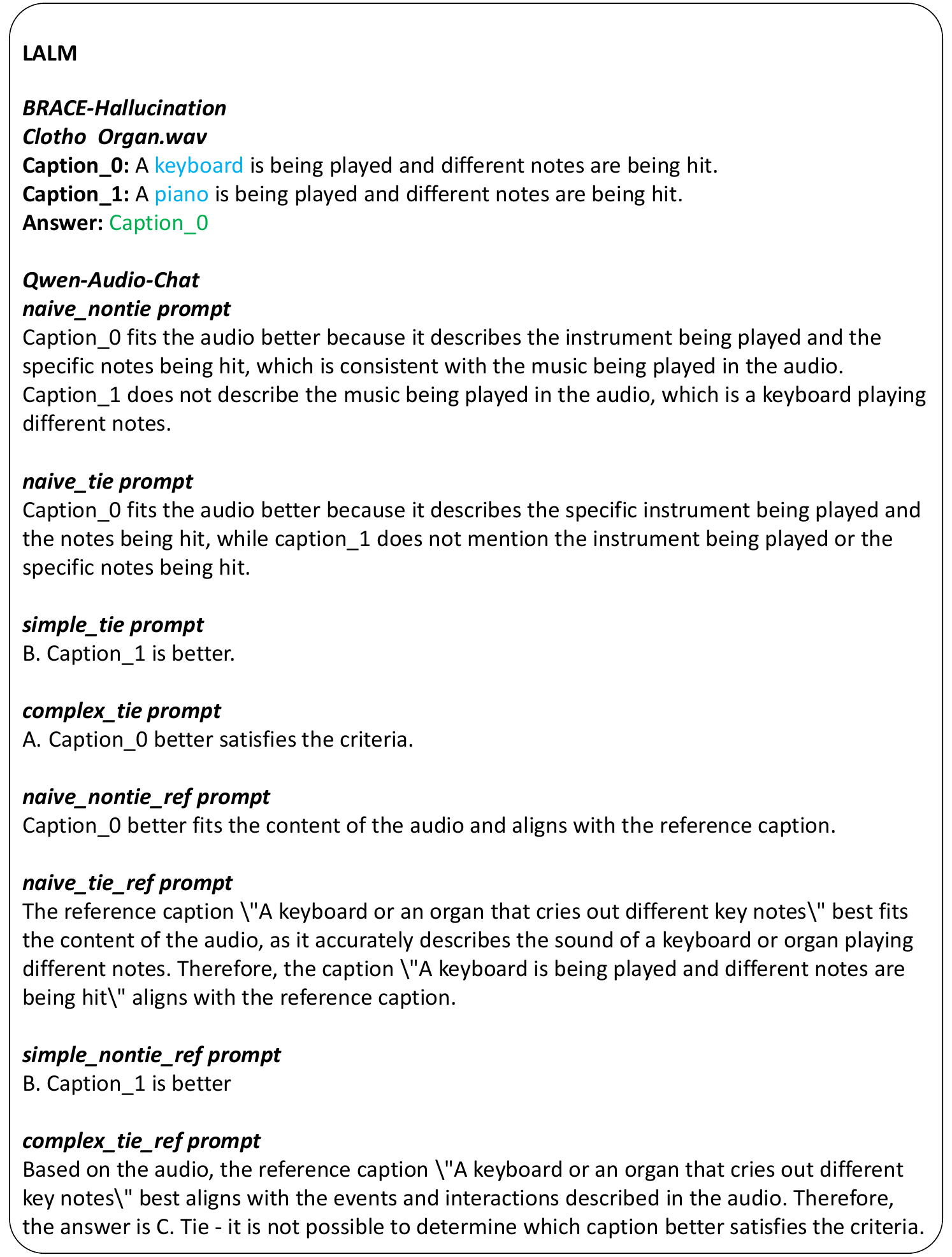}
    \caption{Cases of the LALMs on the BRACE-Hallucination Benchmark}
    \label{figexamples_lalms_appendix_2}
\end{figure}

Figure~\ref{figexamples_lalms_appendix_2} illustrates the performance of Qwen-Audio-Chat, a state-of-the-art LALM on the BRACE benchmark, when responding to the same question under different selected prompts. The results reveal that prompt formulation has a substantial impact on model behavior. 

Notably, when reference captions are included within the prompt, the model sometimes mistakenly selects the reference as the optimal answer, misinterpreting the task objective. This behavior persists even under complex prompts that explicitly and rigorously define the evaluation criteria, suggesting that LALMs remain susceptible to prompt-induced biases and instruction misinterpretation. 

\clearpage
\section{Prompts}
\label{appendix:prompts}
\begin{figure}[h]
    \vspace{4mm}
    \centering
    \includegraphics[width=\textwidth]{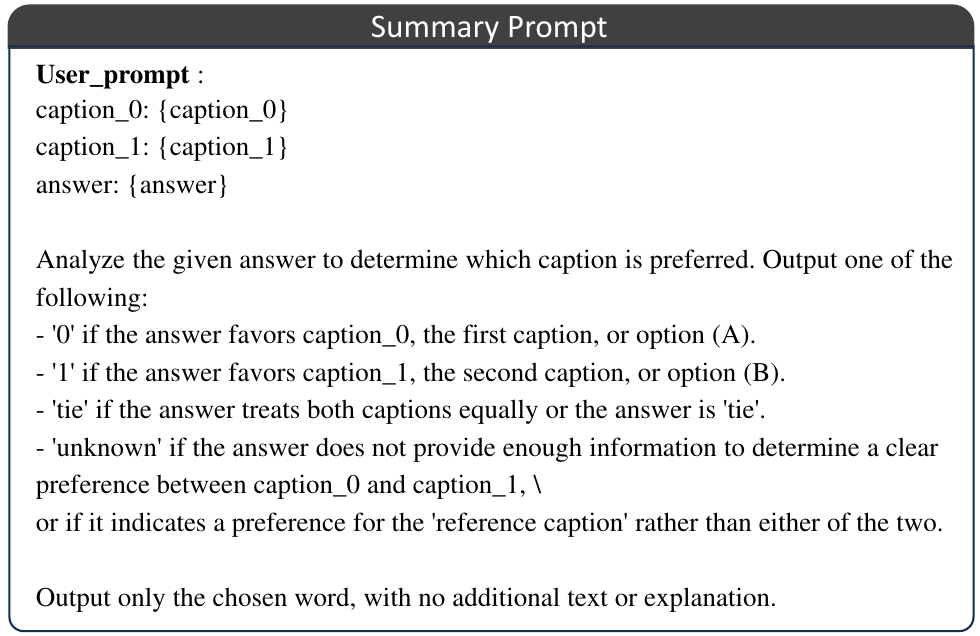}
    \caption{Summary Prompt}
    \label{fig: summary_prompt}
\end{figure}

\begin{figure}[h]
    \centering
    \includegraphics[width=\textwidth]{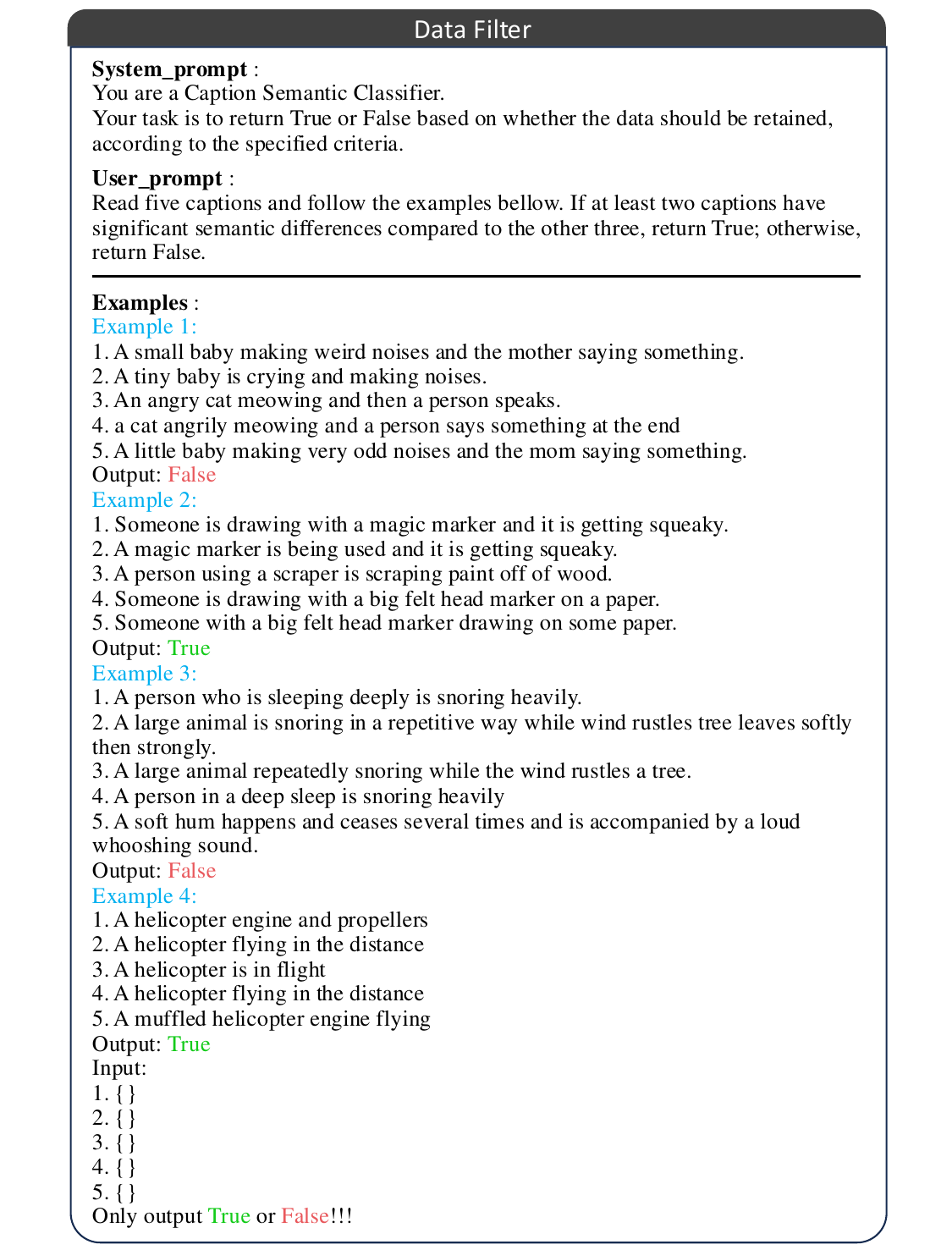}
    \caption{Data Filter Prompt}
    \label{fig: data_filter_prompt}
\end{figure}

\begin{figure}[h]
    \centering
    \includegraphics[width=\textwidth]{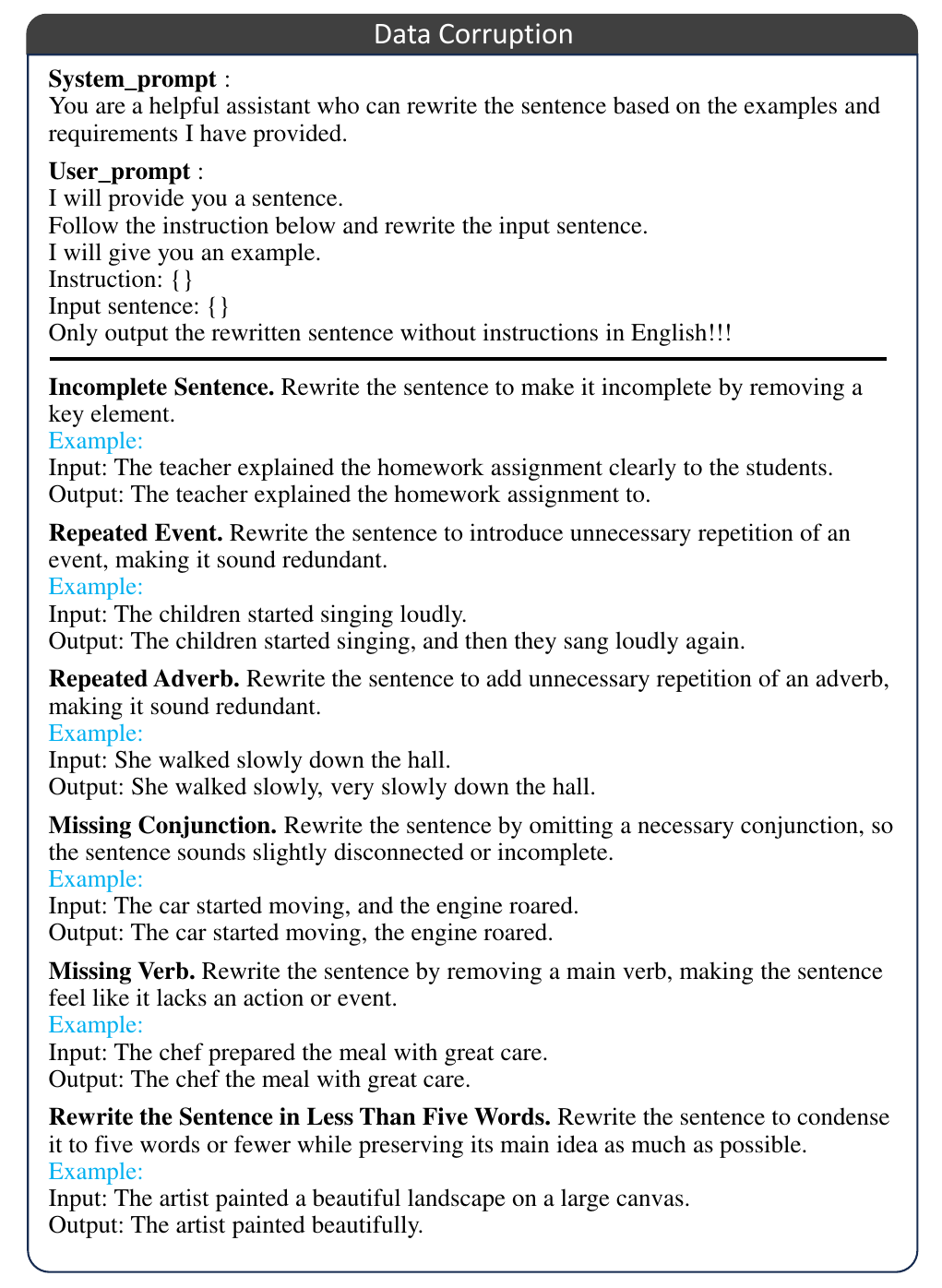}
    \caption{Data Corruption Prompt}
    \label{fig: data_corruption_prompt}
\end{figure}

\begin{figure}[h]
    \centering
    \includegraphics[width=\textwidth]{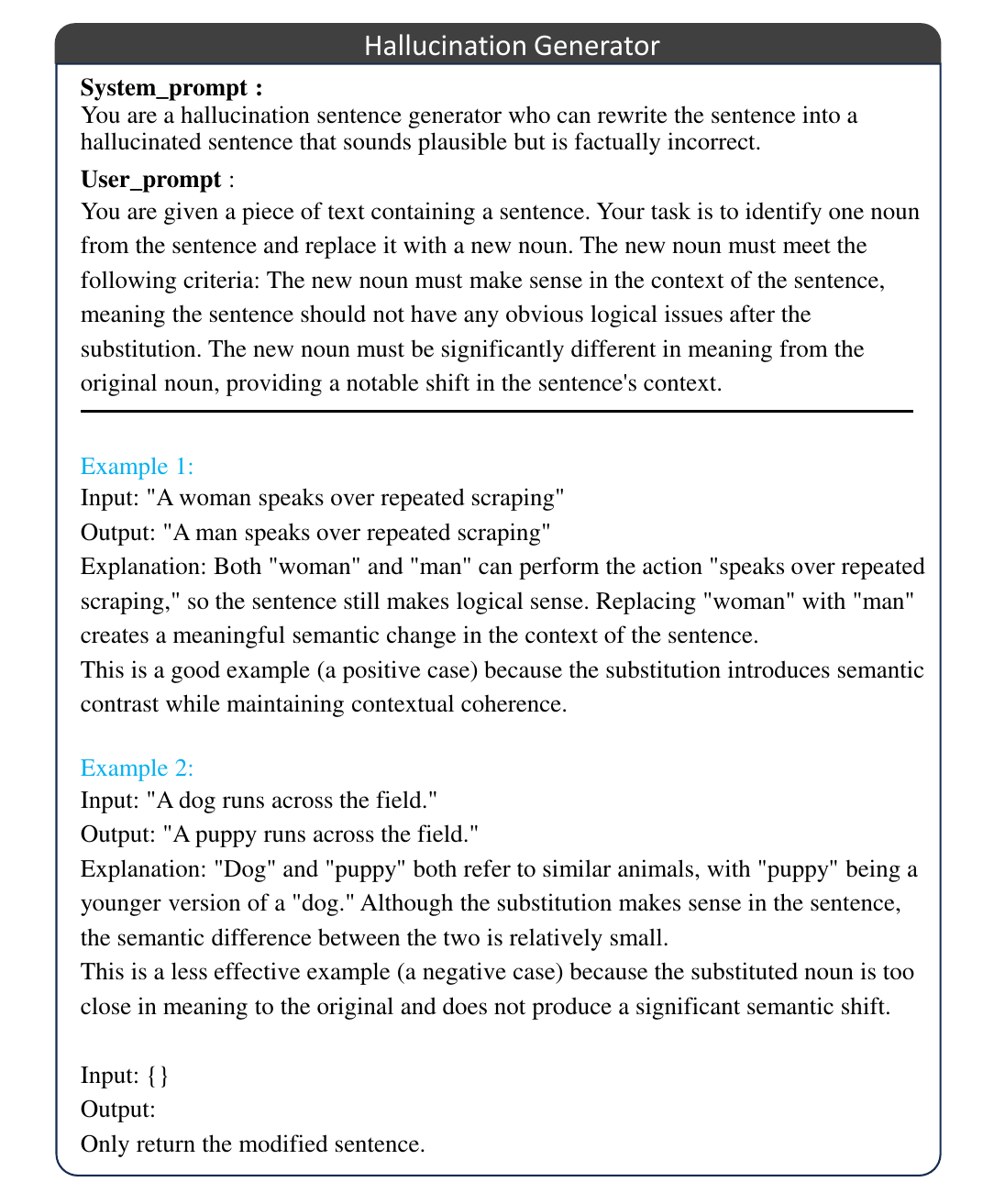}
    \caption{Hallucination Generation Prompt}
    \label{fig: hallucination_generator_prompt}
\end{figure}

\begin{figure}[h]
    \centering
    \includegraphics[width=\textwidth]{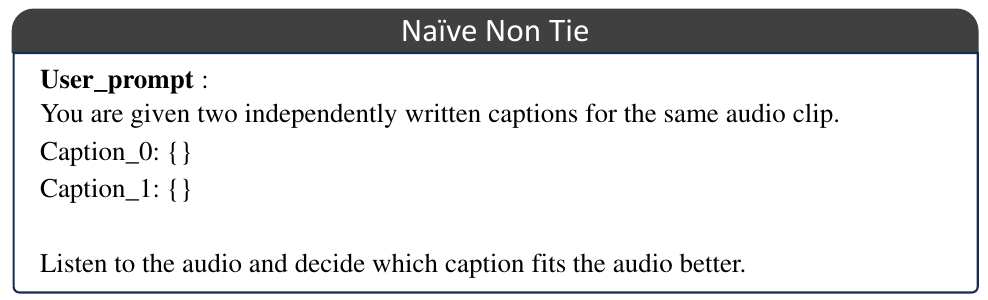}
    \caption{Naive Non Tie Prompt}
    \label{fig: naive_nontie_prompt}
\end{figure}

\begin{figure}[h]
    \centering
    \includegraphics[width=\textwidth]{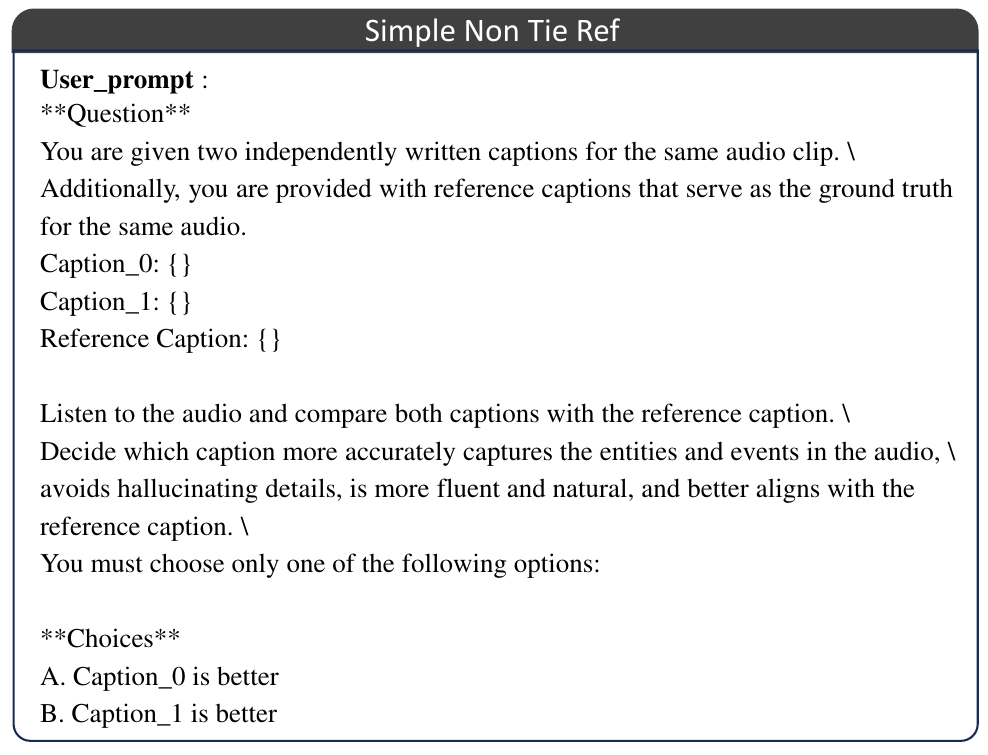}
    \caption{Simple Non Tie with ref Prompt}
    \label{fig: simple_nontie_ref_prompt}
\end{figure}

\begin{figure}[h]
    \centering
    \includegraphics[width=\textwidth]{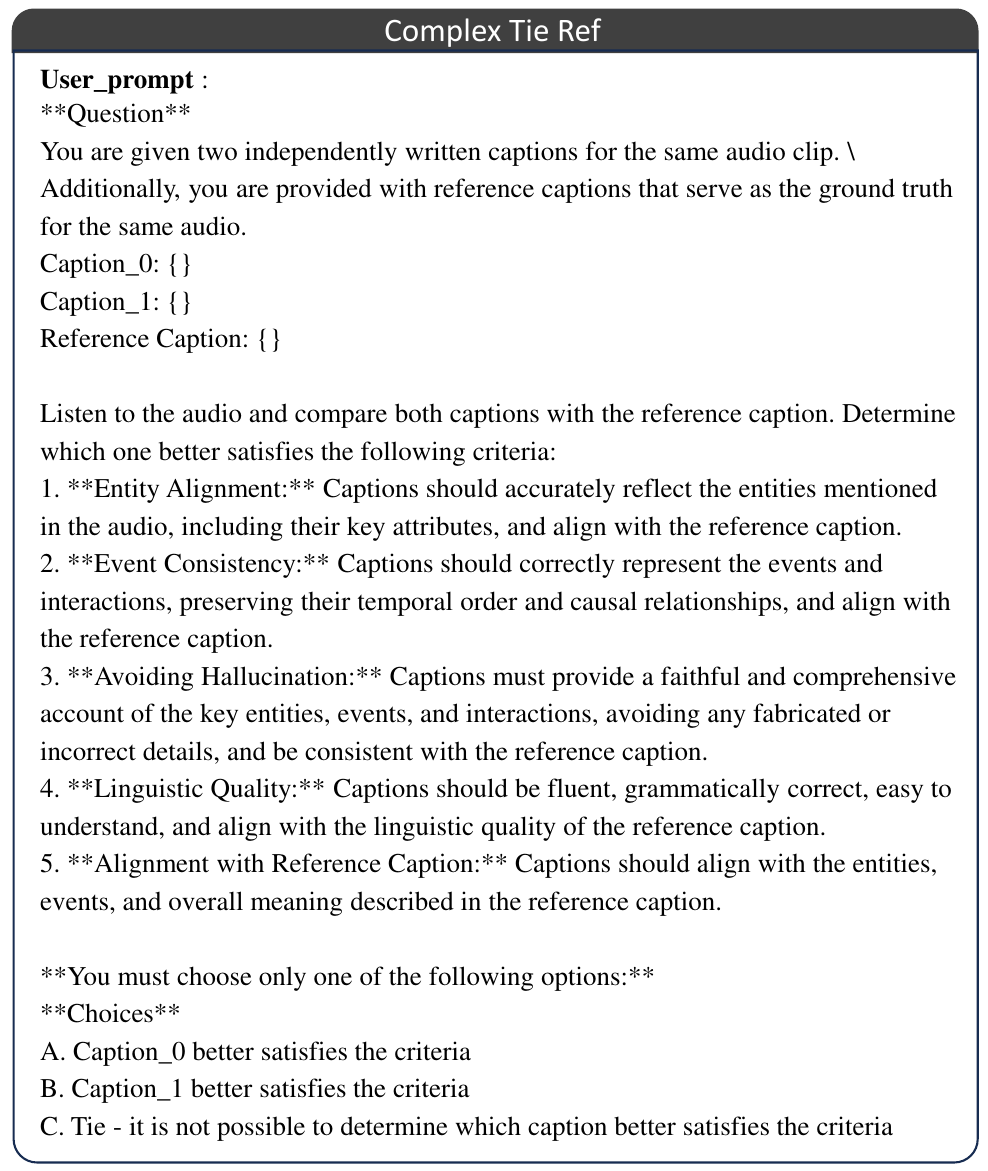}
    \caption{Complex Tie with Ref Prompt}
    \label{fig: complex_tie_ref_prompt}
\end{figure}

In this section, we present the prompts used in our study. 
\begin{itemize}
\item Figure~\ref{fig: summary_prompt} shows the prompt used to process LALM outputs with a large language model. The primary goal is to have the language model evaluate whether the choice made by the LALM is appropriate based on its output. 
\item Figure~\ref{fig: data_filter_prompt} illustrates our filtering process applied to the evaluation sets of AudioCaps and Clotho using Qwen2.5-7B-Instruct. We adopt a few-shot approach, where the captions associated with each audio clip are provided as input to the model. The model is then prompted to determine whether the captions consistently describe the same audio scene. 
\item Figure~\ref{fig: data_corruption_prompt} illustrates the process used to corrupt machine-generated captions. We first prompt the model to shorten the original caption, and then introduce fluency errors, including Incomplete Sentences, Repeated Events, Repeated Adverbs, Missing Conjunctions, and Missing Verbs.
\item Figure~\ref{fig: hallucination_generator_prompt} presents the prompt used to generate hallucinated data based on human-annotated captions. We provide explicit generation rules, along with illustrative examples and corresponding explanations, to guide the model in understanding and producing the desired hallucinated content. 
\item Figure~\ref{fig: naive_nontie_prompt}, Figure~\ref{fig: simple_nontie_ref_prompt} and Figure\ref{fig: complex_tie_ref_prompt} are examples of prompts used as input to the LALMs. We design three levels of prompting: \textit{naive}, \textit{simple}, and \textit{complex}. The \textit{naive} prompt directly instructs the model to select the caption that best aligns with the audio. The \textit{simple} prompt highlights key considerations the model should take into account during the selection process. The \textit{complex} prompt provides detailed, rule-based guidance to ensure consistent and reliable decision-making. In addition, we construct two variants of each prompt type to test whether the model is capable of outputting a "tie" when no caption aligns well with the audio. We also provide an additional set of prompts that include reference captions as part of the input. More prompts can be seen in our github repository.
\end{itemize}

\clearpage

\clearpage
\section{Limitations} \label{appendix:limitations}
The construction of the BRACE benchmark is restricted by open source dataset. The limited diversity of many existing open source datasets can restrict the ability of the benchmark to reflect real-world scenarios, leading to models that perform well in benchmark settings, but fail to generalize in different languages, cultures, and acoustic environments. 
Future work should focus on expanding the diversity of datasets, incorporating multilingual, cross-cultural and acoustically varied samples to enhance the benchmark’s representativeness and the model's real-world robustness.

\section{Ethics Statement} \label{appendix:ethics statement}
Our benchmark datasets utilize human-annotated captions and synthetic data generated by LALMs and LLMs based on existing open source datasets and strict rules. However, certain data may involve content where human captions from original datasets or machine-generated captions may exibit inherent biases. We recommend that future use of this benchmark undergo an additional round of manual review.

\section{Broader Impacts} \label{appendix:broader impact}
Our work does not have a direct negative impact on society. However, preventing misuse of open source audio-caption dataset for data privacy or large-scale generation of harmful content remains an important issue worthy of attention.

\end{document}